\documentclass[aps,prl,reprint,superscriptaddress,nofootinbib]{revtex4-2}
\usepackage{tikz, braket}
\usetikzlibrary{decorations.markings} 
\usetikzlibrary{decorations.pathreplacing, calc, arrows.meta}

\usepackage{amsmath,amssymb,mathtools,bm}

\usepackage{graphicx}
\usepackage[colorlinks=true,citecolor=blue,linkcolor=blue,urlcolor=blue]{hyperref}
\usepackage{comment,ulem}
\newcommand{\Tr}{\operatorname{Tr}}

\newcommand{\Ddeh}{\mathcal{D}_{\rm DEH}}

\newcommand{\Or}{\mathcal{O}}
\newcommand{\g}{\mathfrak{g}}
\newcommand{\G}{\mathfrak{G}}
\newcommand{\uu}{\mathcal{U}}
\newcommand{\stab}{\operatorname{Stab}}

\newtheorem{theorem}{Theorem}
\newtheorem{corollary}{Corollary}
\newtheorem{definition}{Definition}

\makeatletter
\newcommand*\bigcdot{\mathpalette\bigcdot@{.5}}
\newcommand*\bigcdot@[2]{\mathbin{\vcenter{\hbox{\scalebox{#2}{$\m@th#1\bullet$}}}}}
\makeatother

\usepackage{pgfplots}
\usepackage{subcaption}
\usepackage{ragged2e}

\usetikzlibrary{plotmarks}
\pgfplotsset{compat=1.18}

\newsavebox{\fiveSpinInsetBox}

\begin{document}

\title{Noether Symmetries Generate Deterministic Energy Harvesting Protocols}

\author{Ali Akil}
\affiliation{Department of Physics, City University of Hong Kong, Tat Chee Avenue, Kowloon, Hong Kong SAR}
\author{M. Hamed Mohammady}
\affiliation{RCQI, Institute of Physics, Slovak Academy of Sciences, D\'ubravsk\'a cesta 9, Bratislava 84511, Slovakia.}
\author{Zihan Wang}
\affiliation{Department of Physics, City University of Hong Kong, Tat Chee Avenue, Kowloon, Hong Kong SAR}
\author{Oscar Dahlsten}
\affiliation{Department of Physics, City University of Hong Kong, Tat Chee Avenue, Kowloon, Hong Kong SAR}
\date{\today}

\begin{abstract} We consider the general principles for when deterministic energy harvesting (DEH) is possible. DEH means absorbing energy from a fluctuating source without entropy being absorbed. 
We show that the symmetry structure of the
source--harvester dynamics gives a general route beyond existing examples to identify DEH capable source states.  Any
continuous symmetry with a conserved Noether charge induces a source-side orbit
of states that all implement the same deterministic harvester transition,
provided the harvester boundary states are symmetry invariant.  Consequently,
one DEH capable source state with nonzero asymmetry  between charge sectors can
generate infinitely many simultaneously DEH capable source states.  A Jaynes-Cummings model, a three-spin
XX chain, and an SU(2) model illustrate the construction. 
We further extend Noether's theorem to generalised probabilistic theories and thereby generalise our main result. We further establish an asymmetry bound for harvesting, showing that source asymmetry cannot increase on average, while DEH saturates the bound exactly preserving asymmetry.
\end{abstract}

\maketitle

\noindent{\bf\textit{Introduction.---}}
The study of how to extract energy deterministically from fluctuating sources has both practical and foundational importance. In the foundations of thermodynamics, it is common to treat randomness as undesirable, consistent with higher entropy having less thermodynamic value, see e.g.~\cite{Allahverdyan2004,AlickiFannes2013,Binder2015,Ferraro2018,CampaioliRMP2024}. In practical energy harvesting scenarios, the load typically needs a well-defined, non-random, input. For example, a battery typically requires a well-defined voltage in order to charge. Intermediate circuits are therefore applied between the source and load to condition the power. The energy loss of those intermediate circuits is significant e.g.\ in low-voltage regimes~\cite{horowitz1989art,Szarka2012,surender2021rectenna}.  

Deterministic Energy Harvesting (DEH) protocols recently appeared as an alternative approach to power conditioning~\cite{MengPRA2025,wang2026nonunique}.  
A harvester system undergoes the same state transition from a given initial low energy state to a higher energy state at a fixed time $\tau$ for many different source preparations. Only the harvester's intermediate state trajectory depends on the initial source state. Thus at time $\tau$, energy has been absorbed without the source randomness becoming absorbed. There is no feedback or diodes~\cite{MengPRA2025}. A paradigmatic example is a two-level quantum harvester that can absorb energy deterministically from a 1-D force profile $A\cos(\omega t+\phi)\bf{\hat{x}}$ with unknown phase  $\phi$, via quantum Rabi oscillation~\cite{MengPRA2025}. In contrast, standard harvesters, which can be modelled as harmonic oscillators, can provably not avoid picking up $\phi$~\cite{MengPRA2025}.  DEH has also been investigated in a fully quantum Jaynes--Cummings
description~\cite{JaynesCummings1963,ShoreKnight1993}, with the non-uniqueness  of source state decompositions being associated with large sets of
 source states for which DEH is possible~\cite{wang2026nonunique}.  DEH is nevertheless not possible from thermal source states, consistent with the 2nd law of thermodynamics~\cite{wang2026nonunique}.

A key challenge in developing DEH further is to understand more generally when non-trivial viable DEH protocols exist. Are DEH-capable sources isolated examples, or can they be understood via a fundamental physical principle?
\begin{figure}[t]
 \resizebox{\linewidth}{!}{%
\begin{tikzpicture}[>=stealth]
    \coordinate (LeftCenter) at (0, 0);
    \coordinate (RightCenter) at (11, 0);

    \draw[thick] (LeftCenter) ellipse (0.8 and 2.5);
    \draw[thick] (RightCenter) ellipse (0.8 and 2.5);
    
    \draw[thick] (0, 2.5) to[bend right=8] (11, 2.5);
    \draw[thick] (0, -2.5) to[bend left=8] (11, -2.5);

    \coordinate (TL) at ({0.8*cos(30)}, {2.5*sin(30)});  
    \coordinate (BL) at ({0.8*cos(-30)}, {2.5*sin(-30)}); 
    \coordinate (TR) at ({11+0.8*cos(150)}, {2.5*sin(150)}); 
    \coordinate (BR) at ({11+0.8*cos(210)}, {2.5*sin(210)}); 

    \tikzset{
        midarrow/.style={
            ultra thick,
            decoration={markings, mark=at position 0.55 with {\arrow{>}}},
            postaction={decorate}
        }
    }

    
    \draw[midarrow] (BL) arc[start angle=-30, end angle=30, x radius=0.8, y radius=2.5] node[midway, left, xshift=-0.05cm] {\Large \(R_s(\theta)\)};
    
    \draw[midarrow] (TL) to[out=-8, in=188] node[midway, below, yshift=-0.05cm] {\Large \(U(t)\)} (TR);

    \draw[midarrow] (BL) to[out=8, in=172] node[midway, above, yshift=0.05cm] {\Large \(U(t)\)} (BR);
    
    \draw[midarrow] (TR) arc[start angle=150, end angle=210, x radius=0.8, y radius=2.5] node[midway, right, xshift=0.05cm] {\Large \(R_s^\dagger(\theta)\)};


    \filldraw (TL) circle (2.5pt);
    \filldraw (BL) circle (2.5pt);
    \filldraw (TR) circle (2.5pt);
    \filldraw (BR) circle (2.5pt);

    \node[above right, xshift=0.05cm, yshift=0.05cm] at (TL) {\Large \(|g\rangle_h \otimes |\psi_0\rangle_s\)};
    \node[below right, xshift=0.05cm, yshift=-0.05cm] at (BL) {\Large \(|g\rangle_h \otimes |\psi_{\theta}\rangle_s\)};
    
    \node[above left, xshift=-0.05cm, yshift=0.05cm] at (TR) {\Large \(|e\rangle_h \otimes |\phi_0\rangle_s\)};
    \node[below left, xshift=-0.05cm, yshift=-0.05cm] at (BR) {\Large \(|e\rangle_h \otimes |\phi'\rangle_s\)};

\node[above left, xshift=1.3cm, yshift=0.05cm] at (TR) {\Large \(\Or_s\)};

\end{tikzpicture}}
  \captionsetup{
        justification=justified,
        singlelinecheck=false,
        skip=7pt
    }
    \caption{\justifying
    {\bf Noether symmetry $R$ generates orbits over different source states that jointly achieve DEH.} The points in the figure depict source-harvester states, which undergo evolutions $R$ or $U$.  
  $R=R_h\otimes R_s$ has two properties (i) Noether symmetry: $R^\dagger U R=U$, (ii) Leaves harvester energy eigenstates invariant: $R\ket{g/e}_h\ket{\psi}_s=\ket{g/e}_h \ket{\psi'}_s$. Suppose $\ket{\psi_0}_s$ achieves DEH such that $U\ket{g}_h\ket{\psi_0}_s=\ket{e}_h\ket{\phi}_s$. As depicted, if another initial source state  differs only by $R$, and is in that sense on the same orbit \(\Or_s\) as $\ket{\psi_0}_s$, after $U$ being applied the harvester is also then in state $\ket{e}_h$:  $UR\ket{g}_h\ket{\psi_0}_s=RU\ket{g}_h\ket{\psi_0}_s=\ket{e}_h\ket{\phi'}_s$. }
  \label{fig:schematic}
\end{figure}
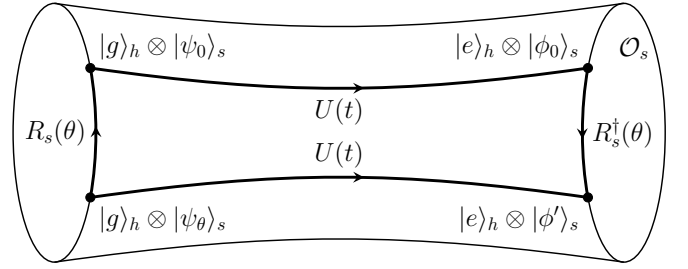
In this paper we show that Noether's theorem~\cite{noether1918invariante} provides such a general generating principle. Originally connecting continuous symmetries with conserved quantities in classical mechanics, Noether's theorem has been  extended to the quantum regime~\cite{MarvianSpekkens2014, Cirstoiu2020}, constituting a fundamental cornerstone of the resource theory of asymmetry, with applications in quantum information~\cite{Marvian2022, Lipka-Bartosik2023a},   quantum measurement~\cite{Kuramochi2022,Mohammady2021a}, quantum reference frames~\cite{Bartlett2007,Loveridge2017a}, and quantum thermodynamics~\cite{LostaglioPRX2015,Marvian2020,Gour2022, YungerHalpern2015,Guryanova2015,Hinds-mingo2018,Majidy2023,Scandi2026}, as well as particle physics~\cite{Weinbergbook}. We here show an additional use of Noether's theorem.

The principle is as follows: a continuous symmetry $R$ of the joint source-harvester dynamics, associated with conserved Noether charge(s), induces  an orbit $\Or$ over source states. This orbit is depicted in Fig.\ref{fig:schematic}. We show that all source states in that $\Or$ lead to the same final state on the harvester at the given time $\tau$. The argument relies on the covariance of $R$ with the joint time evolution $U$, together with the demand that the initial and final states of the harvester are energy eigenstates. In that sense, the deterministic energy transfer is symmetry protected against rotations $R$. One asymmetric seed source state that is DEH capable can generate infinitely many extra sources that are DEH capable. 

The approach allows a general method for identifying a multitude of non-trivial DEH protocols. We illustrate this with several examples. We moreover show that the indirect measurement on the source associated with the harvester eigenstates is invariant under the symmetry and that the source asymmetry is invariant under exact Deterministic Energy Harvesting. Finally, we generalise Noether's theorem  and the associated DEH argument to generalised probabilistic theories, which encompasses quantum, classical and more general dynamics.


\noindent
{\bf\textit{DEH: multiple source states lead to same final harvester state.---}}
Deterministic energy harvesting is naturally formulated as a state-to-state task for the harvester.
\begin{definition}[Source--harvester dynamics]
Let \(h\) denote the harvester and \(s\) the source, with composite Hamiltonian
\begin{equation}
H=H_h+H_s+H_{\mathrm{int}},
\end{equation}
where \(H_h\) and \(H_s\) are the free Hamiltonians and \(H_{\mathrm{int}}\) mediates the energy exchange. For any time $t$, the joint evolution is
\(
U(t)=\exp(-i t H ).
\)
For a source preparation \(\sigma_s\), and at a fixed harvesting time \(t = \tau\), the induced reduced channel on the harvester is
\begin{equation}
\Phi_{\sigma_s}(\rho_h)
=
\Tr_s\!\left[
U(\tau)(\rho_h\otimes\sigma_s)U^\dagger(\tau)
\right].
\label{eq:channel}
\end{equation}
\end{definition}
We only consider cases where the initial and final states have zero interaction energy, to ensure that the harvested energy is solely from the source.
\begin{definition}[Deterministic energy harvesting]
A deterministic harvesting task is specified by two boundary pure states of the harvester, an initial pure ground state \(\rho_h^i\) and a target final pure excited state \(\rho_h^f\). A source state \(\sigma_s\) achieves deterministic energy harvesting for this task if
\begin{equation}
\Phi_{\sigma_s}(\rho_h^i)=\rho_h^f.
\label{eq:deh}
\end{equation}

\end{definition}

The word deterministic refers to the fact that, at the prescribed time \(\tau\), the harvester ends in the exact same pure state \(\rho_h^f\), not merely with the same average energy. Thus no classical or quantum uncertainty in the admissible source preparation is transferred to the final state of the load. This motivates collecting all such source state preparations enabling such a deterministic transition into a single set.

\begin{definition}[DEH-capable source set]
The set of source states jointly implementing the deterministic task
\(\rho_h^i\to\rho_h^f\) is
\begin{equation}
\mathcal D_{\mathrm{DEH}}(\rho_h^i\!\to\!\rho_h^f)
=
\left\{
\sigma_s:
\Phi_{\sigma_s}(\rho_h^i)=\rho_h^f
\right\}.
\end{equation}
\end{definition}

The central question is then how this set $\mathcal D_{\mathrm DEH}$ of task-achieving source states is organized. The Noether construction below shows that, whenever the dynamics has an appropriate continuous symmetry, a single DEH-capable source with asymmetry, which we call a \textit{seed state}, can generate an entire orbit of physically distinct source states lying inside the same set \(\mathcal D_{\mathrm{DEH}}\).

{\noindent\bf
\textit{Noether symmetries.—}}
Continuous symmetries connect conserved charges to the dynamical symmetry used below. Let \(\mathfrak G\) be a connected Lie group with local strongly continuous unitary representations   \(R_x(\mathfrak g)\), \(x=h,s\), and
\(
R(\mathfrak g)=R_h(\mathfrak g)\otimes R_s(\mathfrak g).
\) 
For $\g$ close to the identity element of the group, we may write $R(\mathfrak g)=\exp(i\sum_{\alpha}\theta^{\alpha} Q^\alpha)$, with additive  Hermitian generators 
$Q^\alpha=Q_h^\alpha\otimes I_s+I_h\otimes Q_s^\alpha.$

By Noether's theorem, conservation of these charges, \([H,Q^\alpha]=0\), is equivalent to invariance of the dynamics under the generated continuous symmetry,
\[
[U(t),R(\mathfrak g)]=0
\qquad \forall t, \forall\,\mathfrak g\in\mathfrak G .
\]
This commutation relation is the dynamical symmetry property used below~\cite{noether1918invariante,arnold1989mathematical,dirac1930principles}.

We use the corresponding group actions on operators $A_x$ and maps $\Lambda_x$ :
\begin{align}
\mathfrak g\cdot A_x
&:=R_x(\mathfrak g)A_xR_x^\dagger(\mathfrak g), \nonumber\\
[\mathfrak g\star\Lambda_x](\bigcdot)
&:=R_x(\mathfrak g)\Lambda_x\!\left[
R_x^\dagger(\mathfrak g)(\bigcdot)R_x(\mathfrak g)
\right]R_x^\dagger(\mathfrak g).
\label{eq:G-action}
\end{align}
The symmetry generates the source orbit of states
\begin{equation}
\Or^{\mathfrak{G}}_\sigma=\{\sigma_s^{(\mathfrak g)} = \g\cdot \sigma_s : \mathfrak{g}\in {\mathfrak{G}}\}.
\end{equation}
An operator is invariant if \(\mathfrak g\cdot A_x=A_x\) for all \(\mathfrak g\), and a map is covariant if \(\mathfrak g\star\Lambda_x=\Lambda_x\) for all \(\mathfrak g\). 

\noindent {\bf \textit{Noether theorem argument for deriving DEH source orbits.---}}
We now apply this symmetry structure to the reduced harvester dynamics.
\begin{theorem}
Let \(Q^\alpha=Q_h^\alpha\otimes I_s+I_h\otimes Q_s^\alpha\) be conserved additive Noether charges, \([H,Q^\alpha]=0\), generating the symmetry transformations \(R(\mathfrak{g})=R_h(\mathfrak{g)}\otimes R_s(\mathfrak{g})\).
If the harvester's boundary pure states are locally invariant,
\begin{equation}
  R_h(\mathfrak{g}){\rho_h}^iR_h^\dagger(\mathfrak{g})=\rho_h^i,\quad
  R_h(\mathfrak{g})\rho_h^fR_h^\dagger(\mathfrak{g})=\rho_h^f,
  \label{eq:boundary}
\end{equation}
then every transformed source \(\sigma_s^{(\mathfrak{g})}\) achieves the same deterministic
transition:
\begin{equation}
  \Phi_{\sigma_s^{(\mathfrak{g})}}(\rho_h^i)=\rho_h^f,~ \forall \mathfrak{g} \in \mathfrak{G}.
  \label{eq:noetherdeh}
\end{equation}
Equivalently,
\begin{equation}
  \hspace{-3mm} {\sigma_{\! s}}\in\Ddeh(\rho_h^i\! \! \to \! \rho_{h_{\!f}})
  \Longrightarrow
  \Or^{\mathfrak{G}}_{\sigma_{\! s}} \! \subseteq\Ddeh(\rho_h^i\!\! \to\!\rho_{h_{\!f}}\!),
  \label{eq:orbitclosure}
\end{equation}
where
$
  \Or^{\mathfrak{G}}_\sigma
$ is the group orbit of the source state.
\end{theorem}

{\noindent \bf \textit{Schematic Proof.---}}
Given that $R(\g)$ is a symmetry of the joint dynamics, i.e., $[U(\tau), R(\g)] = 0$, then the unitary channel $\uu(\bigcdot) \coloneq U(\tau) (\bigcdot) U^\dagger(\tau)$ is covariant. Equations \eqref{eq:channel} and \eqref{eq:G-action}, together with covariance of the partial trace, therefore imply that for any source state $\sigma_s$ it holds that $\g \star \Phi_{\sigma_s} = \Phi_{\g\cdot \sigma_s}$. Now assume that the initial harvester state $\rho_h^i$ is invariant, and that for some source seed $\sigma_s$,  the final harvester state $ \rho_h^f = \Phi_{\sigma_s} (\rho_h^i)$ is invariant. It follows that $\rho_h^f = \g \cdot \rho_h^f = [\g\star \Phi_{\sigma_s}] (\rho_h^i) = \Phi_{\g\cdot \sigma_s} (\rho_h^i) $ for all $\g \in \G$, which proves the claim. The details are to be found in the Supplementary Material sec.I.

The symmetry generates a continuous family of physically distinct source states precisely when the source has a nontrivial orbit; this occurs precisely when the source seed $\sigma_s$ is non-invariant, or asymmetric. The physically distinct states in the orbit are labelled by the relative complement 
\(
  \Or^{\mathfrak{G}}_{\sigma_s}\simeq {\mathfrak{G}}\setminus \stab_{\mathfrak{G}}(\sigma_s),
\) 
where \(\stab_{\mathfrak{G}}(\sigma_s)\) is the subgroup leaving \(\sigma_s\) invariant.  Thus
the family is continuous whenever this relative complement has cardinality greater than 1.
Infinitesimally, the orbit is nontrivial if $
  [Q_s^\alpha,\sigma_s]\neq0 $. 


\begin{corollary}
Suppose the harvester charge is compatible with its free Hamiltonian,
\begin{equation}
[Q_h^\alpha,H_h]=0,
\end{equation}
and the harvester boundary states are nondegenerate energy eigenstates of
\(H_h\). Then they are invariant under the symmetry generated by
\(Q_h^\alpha\), and hence Theorem~1 applies.
\end{corollary}
More generally, since \([Q_h^\alpha,H_h]=0\) implies
\([Q_h^\alpha,f(H_h)]=0\), any boundary
state that is a function of \(H_h\), such as a
a thermal state, or the normalized projector onto a whole degenerate energy
shell, is automatically invariant under \(R_h(\mathfrak{g})\). 
In these common cases the
condition in Eq.~\eqref{eq:boundary} is not an additional dynamical
assumption; it follows from the fact that the charging task is insensitive to the
phase or internal coordinate generated by the conserved harvester charge.

\noindent\textit{\bf{Robustness.---}} Theorem 1 assumes an exact Noether symmetry of the joint dynamics and exact invariance of the harvester boundary states. Both assumptions can be relaxed. Suppose that, for the group elements of interest and every joint harvester--source state \(\varrho\),
\[
\left\|
U(\tau)\varrho U(\tau)^\dagger \! \!
-\!
R(\mathfrak g)U(\tau) R^\dagger(\mathfrak g)\,
\varrho\,
R(\mathfrak g)U(\tau)^\dagger R^\dagger(\mathfrak g)
\right\|_1
\leq \epsilon_{\rm cov}.
\]
If the seed transition has error \(\epsilon_{\rm seed}\), and the initial and final harvester states violate symmetry invariance by at most \(\epsilon_i\) and \(\epsilon_f\), respectively, then every transformed source satisfies
\begin{equation}
\left\|
\Phi_{\sigma_s^{(\mathfrak g)}}(\rho_h^i)-\rho_h^f
\right\|_1
\leq
\epsilon_{\rm seed}+\epsilon_{\rm cov}+\epsilon_i+\epsilon_f .
\end{equation}
The derivation is given in the Supplementary Material. Thus, the Noether-generated family is stable under approximate symmetry and approximate DEH. In the exact-symmetry limit, an approximate seed generates an orbit of sources with the same harvesting accuracy.
\\

\noindent {\bf \textit{Example: Number conservation and phase orbits.---}}
For any U(1)-invariant Hamiltonian of a source and harvester, the total excitation number is conserved. The theorem reveals phase-rotated source states with the same deterministic harvester transition. Let
\begin{equation}
  Q= N:=n_h\otimes I_s +I_h\otimes n_s,\qquad [H,N]=0,
\end{equation}
with \(R_h(\theta)=e^{i\theta n_h}\) and
\(R_s(\theta)=e^{i\theta n_s}\).  If the harvester starts and ends in
number-diagonal states, e.g. the ground state $|g\rangle$ and excited state $|e\rangle$ in the Jaynes-Cummings model,  then a source phase rotation $
  \sigma_\theta=e^{i\theta n_s}\sigma e^{-i\theta n_s} $
gives the same harvester transition as \(\sigma\).  In a source number basis,
\begin{equation}
  \sigma_\theta
  =
  \sum_{m,n}e^{i(m-n)\theta}\sigma_{mn}|m\rangle\langle n|.
\end{equation}
The orbit is trivial for number-diagonal \(\sigma\), but nontrivial whenever
\([n_s,\sigma]\neq0\).  A single DEH source with coherence between number
sectors therefore implies a continuous family of phase-distinct sources with
identical deterministic output.

Phase-insensitive harvesting is in that sense a consequence of conserved excitation number rather than an accident of a particular Rabi solution~\cite{JaynesCummings1963,ShoreKnight1993}.  The randomness associated with
the source phase is symmetry-protected from appearing in the harvester
output.

\noindent{\bf \textit{Example: Non-Abelian source orbits.---}}
We now illustrate the theorem with a non-Abelian symmetry. The harvester has two rotational-scalar boundary states, $|g\rangle_h$ and $|e\rangle_h$, defining the pure charging task 
$
|g\rangle_h\longrightarrow |e\rangle_h .
$
We also include an auxiliary spin-one harvester triplet 
${|a,\alpha\rangle_h},{\alpha=x,y,z},$
so the full harvester Hilbert space still carries a nontrivial $SU(2)$ representation.
The source is a five-spin-$1/2$ chain with total spin $\mathbf J_s=\sum_i\mathbf S_i$ and an isotropic Heisenberg Hamiltonian $H_s=\sum_i J_i\,\mathbf S_i\cdot\mathbf S_{i+1}$, where ${\bf S}_i=\frac12(\sigma_i^x,\sigma_i^y,\sigma_i^z)$.
The source and harvester are coupled through rotationally invariant scalar products of vector operators. Defining $T_s^\alpha=\sum_i c_iS_i^\alpha,$ $W_s^\alpha=\sum_i d_iS_i^\alpha$, 
\(
D_g^\alpha=|a,\alpha\rangle\langle g|+|g\rangle\langle a,\alpha|,
\)   and \(
D_e^\alpha=|a,\alpha\rangle\langle e|+|e\rangle\langle a,\alpha|,
\)
the interaction is
\begin{equation}
H_{\rm int}
=
\lambda_g\! \! \sum_{\alpha=x,y,z} \! \! D_g^\alpha\otimes T_s^\alpha
+
\lambda_e\! \! \sum_{\alpha=x,y,z}\! \! D_e^\alpha\otimes W_s^\alpha .
\label{5_spin_hamiltonian}
\end{equation}
Together with the rotationally invariant $H_s$ and harvester Hamiltonian $H_h$, this makes the full Hamiltonian $H=H_h+H_s+H_{\rm int}$ invariant under joint $SU(2)$ rotations, and hence
\begin{equation} \label{eq:su2_conserved}
[H,J_h^\alpha+J_s^\alpha]=0,
\qquad \alpha=x,y,z~,
\end{equation}
which makes $Q^\alpha=J_h^\alpha +J_s^\alpha$ the system's conserved charges.
The local angular momenta need not be conserved separately, so angular momentum can still be exchanged between source and harvester.

We first use a numerical simulation to find a seed state. For a source state $|\psi\rangle_s$, the pure-target fidelity is
$
F_e(\tau,\! \psi)\!=\!
\left\|
A_{eg}(\tau)|\psi\rangle_s
\right\|^2 \!\!\!, 
\, \,
A_{eg}(\tau)\!=\!{}_h\langle e|e^{-iH\tau}|g\rangle_h .
$
We find a DEH seed by maximizing the largest singular value of $A_{eg}(\tau)$ over the time $\tau$. For the parameter set given in the Supplemental Material, the optimum occurs at $\tau=\tau_\star$ and gives a seed $|\psi_\star\rangle_s$ with
\begin{equation}
F_e(\tau_\star,\psi_\star)=1-\epsilon ,
\qquad
\epsilon\ll1 .
\end{equation}
The optimized seed is not rotationally invariant and therefore generates a nontrivial $SU(2)$ orbit. In the total-spin basis it can be written as
$ |\psi_\star\rangle_s
=
\sum_{J,\mu,m} C_{J\mu m}|J,\mu,m\rangle_s,
$
where $J$ and $m$ label total spin and its magnetic component, while $\mu$ labels multiplicities. Under a rotation $R\in SU(2)$, the source becomes
\begin{align}
|\psi_\star\!(R)\rangle_s 
& =
R_s(R)|\psi_\star\rangle_s \\
&= \hspace{-3mm}\sum_{J,\mu,m,m'} \hspace{-3mm}
C_{J\! \mu m}
D^{(J)}_{m'm}(R)
|J,\mu,m'\rangle_s .
\end{align}
Thus, whenever the seed has support in a nonzero-spin sector, the rotation mixes magnetic components within that sector and generates physically distinct many-body source states. 

Unlike the $U(1)$ phase orbit, where symmetry transformations only attach relative phases to number sectors, the non-Abelian $SU(2)$ action changes the amplitudes among states within each multiplet. Nevertheless, all these source states lead to the same final state on the harvester. 
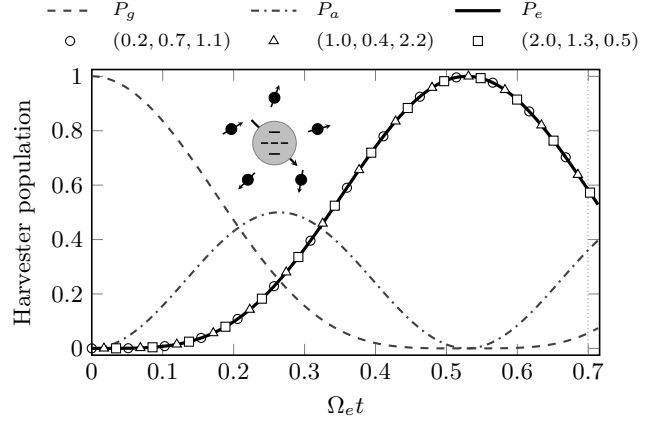
\begin{figure}[htbp!]
    \centering
    \usetikzlibrary{arrows.meta}


\pgfmathsetmacro{\OmegaE}{0.9992837576}
\pgfmathsetmacro{\Tstar}{0.7000333028}
\pgfmathsetmacro{\OmegaETstar}{0.6995319093}


\newcommand{\insetspin}[3]{%
    \begin{scope}[shift={(#1:#2)}, rotate=#3]

        \draw[
            ->,
            line width=2.5pt,
            >=stealth
        ] (0,-0.55) -- (0,0.75);

        \filldraw[black] (0,0) circle (0.3);

    \end{scope}
}

\newcommand{\fiveSpinInset}{%
    \begin{tikzpicture}[
        baseline=(current bounding box.center)
    ]

        \draw[
            ->,
            black,
            line width=3.5pt,
            >=stealth
        ] (135:1.8) -- (-45:1.8);

        \filldraw[
            fill=lightgray,
            draw=gray
        ] (0,0) circle (1.2);

        \draw[
            black,
            line width=2.5pt
        ] (-0.3,-0.6) -- (0.3,-0.6);

        \draw[
            black,
            line width=2.5pt
        ] (-0.75,0) -- (-0.35,0);

        \draw[
            black,
            line width=2.5pt
        ] (-0.20,0) -- (0.20,0);

        \draw[
            black,
            line width=2.5pt
        ] (0.35,0) -- (0.75,0);

        \draw[
            black,
            line width=2.5pt
        ] (-0.3,0.6) -- (0.3,0.6);

        \insetspin{90}{2.5}{-20}
        \insetspin{162}{2.5}{-60}
        \insetspin{234}{2.5}{135}
        \insetspin{306}{2.5}{170}
        \insetspin{18}{2.5}{-75}

    \end{tikzpicture}%
}

\savebox{\fiveSpinInsetBox}{%
    \fiveSpinInset
}


\begin{tikzpicture}

\begin{axis}[
    width=0.96\linewidth,
    height=0.62\linewidth,
    xlabel={$\Omega_e t$},
    ylabel={Harvester population},
    xmin=0,
    xmax=0.716,
    enlarge x limits=false,
    ymin=-0.025,
    ymax=1.025,
    ytick distance=0.2,
    tick align=inside,
    tick label style={font=\small},
    label style={font=\small},
    axis line style={line width=0.6pt},
    legend style={
        at={(0.5,1.03)},
        anchor=south,
        draw=none,
        font=\scriptsize,
        legend columns=3,
        legend cell align=left,
        column sep=5pt,
        row sep=1pt
    }
]

%

\addplot[
    black!45,
    thin,
    densely dotted,
    no marks,
    forget plot
]
coordinates {
    (\OmegaETstar,-0.025)
    (\OmegaETstar,1.025)
};


\addplot[
    black!75,
    thick,
    dashed,
    no marks
]
table[
    x=tau,
    y=Pg
]
{nonabelian_harvester_populations.dat};
\addlegendentry{$P_g$}

\addplot[
    black!75,
    thick,
    dashdotted,
    no marks
]
table[
    x=tau,
    y=Pa
]
{nonabelian_harvester_populations.dat};
\addlegendentry{$P_a$}

\addplot[
    black,
    very thick,
    solid,
    no marks
]
table[
    x=tau,
    y=Pe
]
{nonabelian_harvester_populations.dat};
\addlegendentry{$P_e$}


\addplot[
    only marks,
    black,
    mark=o,
    mark size=1.7pt,
    mark repeat=36,
    mark phase=1
]
table[
    x=tau,
    y=PeB
]
{nonabelian_orbit_populations.dat};
\addlegendentry{$(0.2,0.7,1.1)$}

\addplot[
    only marks,
    black,
    mark=triangle*,
    mark options={fill=white},
    mark size=2.0pt,
    mark repeat=36,
    mark phase=13
]
table[
    x=tau,
    y=PeC
]
{nonabelian_orbit_populations.dat};
\addlegendentry{$(1.0,0.4,2.2)$}

\addplot[
    only marks,
    black,
    mark=square*,
    mark options={fill=white},
    mark size=1.7pt,
    mark repeat=36,
    mark phase=25
]
table[
    x=tau,
    y=PeD
]
{nonabelian_orbit_populations.dat};
\addlegendentry{$(2.0,1.3,0.5)$}


\node[
    anchor=north west,
    inner sep=1pt,
    fill=white
]
at (axis description cs:0.25,0.96)
{%
    \scalebox{0.24}{\usebox{\fiveSpinInsetBox}}%
};

\end{axis}

\end{tikzpicture}
    \captionsetup{
        justification=justified,
        singlelinecheck=false,
        skip=0pt
    }
    \caption{\justifying
    {\bf Non-Abelian DEH protocol with conserved total angular momentum.} A five-level harvester is coupled to a source of five spin-1/2 particles through the interaction Hamiltonian in Eq.~(\ref{5_spin_hamiltonian}). The horizontal axis  shows the dimensionless time $\Omega_e\ t$, where $\Omega_e$ is the harvester transition frequency between $|g\rangle_h$ and $|e\rangle_h$; equivalently, $E_e-E_g = \hbar\Omega_e$. The curves show the ground-state, auxiliary-level, and excited-state
    populations $P_g$, $P_a$, and $P_e$, respectively. The solid $P_e$
    curve corresponds to the unrotated source state
    $(\alpha,\beta,\gamma)=(0,0,0)$, while the circles, triangles, and
    squares represent three globally $\mathrm{SU}(2)$-rotated source
    states with Euler angles specified in the legend. Their coincidence
    demonstrates the invariance of the excitation dynamics along the
    symmetry orbit. The vertical dotted line indicates the optimal
    transfer time $\tau_\star$. }
    \label{fig:nonabelian_dynamics}
\end{figure}
The entire $SU(2)$ orbit implements the same factorized harvester transition,
\begin{equation}
U(\tau_\star)|g\rangle_h|\psi_\star(R)\rangle_s
\simeq
|e\rangle_h|\eta_\star( R)\rangle_s ,
\end{equation}
where $ |\eta_\star( R)\rangle_s
=
R_s(R)|\eta_\star\rangle_s .$
Note that the auxiliary harvester triplet generally participates during the evolution, but it has no residual population or entropy at the harvesting time when the transition is exact. The result is a continuous orbit of distinct five-spin source states producing the same pure harvester transition.

\noindent {\bf\textit{Symmetry constrains harvester's information gain.---}}
The symmetry constrains what information about the source can be accessed through the harvester. Consider the case where, after the interaction between source and harvester, the harvester is measured by an invariant observable \(\{\Pi_k\}\). For any harvester preparation $\rho_h$, the conditional transformations of the source are described by the quantum instrument
\begin{equation}
\Lambda_\rho^k(\bigcdot)
=
\Tr_h\!\left[(\Pi_k\!\otimes\! I_s)U(t)
(\rho_h\!\otimes\!\bigcdot)U^\dagger(t)\right],
\end{equation}
which induces the source POVM \(\{M_k =(\Lambda_\rho^k)^\dagger(I_s) \}\)~\cite{Busch2016a}. Covariance of \(U(t)\), together with invariance of \(\rho_h\) and \(\Pi_k\), implies that each \(\Lambda_\rho^k\) is covariant and each \(M_k\) is invariant~\cite{Keyl1999,Hokkyo2026}. Consequently,
\begin{equation}
\Tr[M_k\sigma^{(\mathfrak g)}]
=
\Tr[M_k\sigma]
\qquad \forall\,\mathfrak g,k ,
\end{equation}
so the harvester cannot distinguish different members of the orbit. In this operational sense, \(R_s(\mathfrak g)\) plays the role of a local gauge transformation on the source: symmetry-invariant harvester observables are insensitive to it, analogous to gauge-invariant observables in gauge theories~\cite{Wilson1974,Banuls2020}.

\noindent{\bf \textit{Source asymmetry invariance.---}}
Covariance also constrains the asymmetry remaining in the source after harvesting. Let \(\mathcal I\) be an asymmetry monotone, i.e., a measure of the source's symmetry-breaking resource, and suppose it obeys selective monotonicity under covariant operations~\cite{Takagi2018}, as do, for example, measures based on quantum Fisher information or Wigner--Yanase skew information. Writing \(p_k=\Tr[\Lambda_\rho^k(\sigma)]\) and \(\sigma_k=\Lambda_\rho^k(\sigma)/p_k\), covariance of the instrument gives
\begin{equation}
\mathcal I(\sigma)\geq\sum_k p_k\,\mathcal I(\sigma_k).
\end{equation}
Thus a general harvesting process cannot increase source asymmetry on average, although it may increase for an individual outcome.

Exact DEH strengthens this inequality to an equality. For the pure boundary states considered here, let \(A_{eg}(\tau)={}_h\langle e|U(\tau)|g\rangle_h\), so that the final source state is \(\sigma'=A_{eg}(\tau)\sigma A_{eg}(\tau)^\dagger\). Exact DEH implies \(A_{eg}(\tau)^\dagger A_{eg}(\tau)=I_s\) on \(\operatorname{supp}\sigma\), and hence \(A_{eg}(\tau)\) acts isometrically on the DEH support, preserving the source entropy. Moreover, symmetry of \(U(\tau)\) and invariance of the harvester boundary states imply
\begin{equation}
A_{eg}(\tau)R_s(\mathfrak g)
=
e^{i\vartheta(\mathfrak g)}R_s(\mathfrak g)A_{eg}(\tau) ,
\end{equation}
so \(A_{eg}(\tau)\) may transfer charge without commuting with the conserved source charge, while the induced transformation \(A_{eg}(\tau)(\,\cdot\,)A_{eg}(\tau)^\dagger\) remains covariant. As shown in the Supplemental Material, the isometry can be extended to covariant channels implementing both \(\sigma\to\sigma'\) and \(\sigma'\to\sigma\). Monotonicity in both directions therefore yields
\begin{equation}
\mathcal I(\sigma')=\mathcal I(\sigma).
\end{equation}
Thus exact DEH consumes source energy while preserving its entropy and asymmetry; away from the exact deterministic limit, only the one-sided monotonicity bound remains in general.

\noindent{\bf \textit{Generalisation beyond quantum theory.—}}
The orbit construction extends to generalised probabilistic theories admitting a phase-space representation. Let $f(z_h,z_s,t)$ denote the joint state and consider dynamics generated by~\cite{jiang2024unification, jiang2024framework} $\partial_t f=\{f,H\}_K\equiv\mathcal L_{H,K} f,$ and 
$
\{A,B\}_K:=\int d\kappa\,K(\kappa)\,
A\sin\!\left(\frac{\kappa}{2}\overleftrightarrow{\Lambda}\right)B,
\quad
\overleftrightarrow{\Lambda}=\overleftrightarrow{\Lambda}_h+\overleftrightarrow{\Lambda}_s ,
$
where the kernel $K(\kappa)$ specifies the dynamical theory; appropriate choices recover, in particular, classical Poisson and quantum Moyal dynamics~\cite{wen2026generalised, jiang2024unification, jiang2024framework, plavala2022operational, plavala2023general}. We generalise Noether's theorem to that dynamics (see Supplementary material). Writing $\mathcal T_{\tau,K}=e^{\tau\mathcal L_{H,K}}$ for the joint evolution and $\mathcal M_s$ for source marginalisation, the reduced harvester dynamics is then
\begin{equation}
\Phi^\tau_{f_s,K}[f_h]:=\mathcal M_s \circ \mathcal T_{\tau,K}[f_hf_s].
\end{equation}
Let $R_h(g)$ and $R_s(g)$ be reversible local symmetry actions, with $R_{hs}(g)=R_h(g)\boxtimes R_s(g)$. If $\mathcal M_s\circ R_{hs}(g)=R_h(g)\circ \mathcal M_s$ and $\mathcal T_{\tau,K}\circ R_{hs}(g)=R_{hs}(g)\circ \mathcal T_{\tau,K}$, then
\begin{equation}
\Phi^\tau_{R_s(g)[f_s], K}[f_h]
=
R_h(g)\Phi^\tau_{f_s,K}[R_h(g)^{-1}[f_h]].
\end{equation}
Hence, if the harvester boundary states are symmetry invariant, the entire source orbit $R_s(g)[f_s]$ implements the same deterministic transition. Thus the Noether--DEH orbit construction applies across classical, quantum, and more general phase-space theories. 

As an explicit classical realisation, we consider a charged rotating sphere with angular momentum $\mathbf L_h$, coupled autonomously to a harmonic-oscillator source with scaled canonical quadratures $X$ and $P$.  The joint dynamics possesses the conserved Noether charge $Q=L^z_h+J_s$, where $L_h^z$ is the $z$ component of the harvester angular momentum and $J_s=(X^2+P^2)/2$ is the source-oscillator action. The resulting phase orbit provides a classical DEH family: all source states with the same initial action and arbitrary phase drive the harvester through the same boundary-state transition at the same time while transferring the same energy; see the Supplemental Material.
\\
\noindent{\bf\textit{Summary and Outlook.---}}We have shown that a continuous symmetry whose action leaves the harvester boundary states invariant generates symmetry-related source states realizing the same deterministic harvesting task. Noether symmetries therefore provide a general organizing principle for families of DEH-capable sources, beyond the specific constructions known previously.

The realisation that symmetry protects deterministic harvesting in this manner will help the field develop, with potential foundational and technological impacts. Fundamentally, DEH provides an elementary protocol that may alter the resource hierarchy of energy sources.  Technologically, it provides a route to make ideas from quantum energetics useful for energy harvesting and power conditioning. We expect that the general results here concerning how to construct DEH protocols will aid in identifying multiple DEH protocols, including in electrical circuits.


\begin{acknowledgments} 
\noindent{\bf \textit{Acknowledgments.---}}We thank Alessio Serafini, Hui Tsz Hin and Yu Wing Chi for fruitful discussions.  This work was supported by The City University of Hong Kong (Project No. 9610623) and Research Grants Council (RGC) General Research Fund  (CityU 11300125). M.H.M acknowledges funding provided by the IMPULZ program of the Slovak Academy of Sciences under the Agreement on the Provision of Funds No. IM-2023-79 (OPQUT).

\noindent{\bf \textit{Data Availability.---}}The codes and data that support the findings and reproduce the results of this study are openly available in the GitHub repository~\cite{MyGitHubData}.
\end{acknowledgments}

\bibliographystyle{apsrev4-2}
\bibliography{refs}

\end{document}


\title{Supplemental Material for ``Noether Symmetries Generate Deterministic Energy-Harvesting Protocols''}

\author{Ali Akil}
\affiliation{Department of Physics, City University of Hong Kong, Tat Chee Avenue, Kowloon, Hong Kong SAR}
\author{M. Hamed Mohammady}
\affiliation{RCQI, Institute of Physics, Slovak Academy of Sciences, D\'ubravsk\'a cesta 9, Bratislava 84511, Slovakia.}

\author{Zihan Wang}
\affiliation{Department of Physics, City University of Hong Kong, Tat Chee Avenue, Kowloon, Hong Kong SAR}
\author{Oscar Dahlsten}
\affiliation{Department of Physics, City University of Hong Kong, Tat Chee Avenue, Kowloon, Hong Kong SAR}

\date{\today}
\maketitle

\tableofcontents

\section{Noether Theorem}
\noindent
{\bf\textit{Noether symmetries.—}}
Continuous symmetries provide a natural way to constrain dynamical evolution.
In Hamiltonian mechanics, a conserved quantity $Q$ satisfies
$\dot Q=\{Q,H\}=0$ classically, while in quantum mechanics a conserved
observable satisfies $[H,Q]=0$~\cite{noether1918invariante,
arnold1989mathematical,dirac1930principles}.
More generally, let $\mathfrak G$ be a connected Lie group represented on the
system by unitaries $R(\mathfrak g)$, with Hermitian generators $Q^\alpha$.
If $[H,Q^\alpha]=0\qquad \forall a$ ,
then the full time evolution is invariant under the symmetry, $[U(t),R(\mathfrak g)]=0
\qquad
\forall\,\mathfrak g\in\mathfrak G . $
This commutation relation is the dynamical symmetry property used below.

Let $\G$ be a group, that is identified with the symmetry of interest. For each system $x= h, s, h+s$ let $R_x : \G \ni \g \mapsto R_x(\g)$ be a strongly continuous unitary representation of $\G$ on the Hilbert space  of  $x$, such that the unitary representation on the total system of harvester-plus-source reads $R(\g) \equiv R_{h+s}(\g) = R_h(\g) \otimes R_s(\g)$.   For each system $x$, the group action on operators  and operations (completely positive maps) reads
\begin{align}\label{eq:G-action}
\mathfrak g\cdot A_x
&:=R_x(\mathfrak g)A_xR_x^\dagger(\mathfrak g), \nonumber\\
[\mathfrak g\star\Lambda_x](\,\bigcdot\,)
&:=R_x(\mathfrak g)
\Lambda_x\!\left(
R_x^\dagger(\mathfrak g)(\,\bigcdot\,)R_x(\mathfrak g)
\right)
R_x^\dagger(\mathfrak g).
\end{align}
An operator is invariant if $\mathfrak g\cdot A_x=A_x$, and an operation 
is covariant if $\mathfrak g\star\Lambda_x=\Lambda_x$, for all
$\mathfrak g\in\mathfrak G$. In particular, a source state generates the
orbit $\sigma_s^{(\mathfrak g)}:=\g\cdot \sigma_s = R_s(\mathfrak g)\sigma_s
R_s^\dagger(\mathfrak g)$.  

$\G$ is a (strong) symmetry of the global dynamics if $[ U(\tau), R(\g)] = 0$ for all $\g\in \G$ and all $\tau$, which implies that the unitary channel $\uu(\bigcdot) \coloneq U(\tau) (\bigcdot) U^\dagger(\tau)$ is covariant for all $\tau$. Now assume that  $\G$ is an $N$-dimensional connected Lie group, representing a continuous symmetry of the dynamics. In such a case, for each system $x$ there are $N$ linearly independent Hermitian generators $\{Q_x^\alpha\}_{\alpha=1}^N$, with the total generators being additive, i.e., $Q^\alpha \equiv Q_{h+s}^\alpha \coloneq  Q_h^\alpha \otimes I_s + I_h \otimes Q_s^\alpha$,  such that for all $\g \in G$ there is a finite number $K$ such that   $R_x(\g) = \prod_{k=1}^K e^{i \sum_\alpha \theta_k^\alpha(\g) Q^\alpha_x}$, where $\theta^\alpha_k(\g)$ are real numbers determined by $\g$. If $\G$ is compact then $K=1$, i.e.,  $R_x(\g) = e^{i \sum_\alpha \theta^\alpha(\g) Q^\alpha_x}$ for all $\g\in \G$.   By Noether's theorem, it follows that $\G$ is a continuous symmetry of the global dynamics if and only if each total generator $Q^\alpha$ is a conserved quantity, i.e., $[U(\tau), Q^\alpha] = 0$ for all $\tau$. 

\section{Proof of the Noether-DEH theorem}
\label{supsec:Noether-DEH}
Let \(\G\) be a group represented on the harvester and source by unitaries
\(R_h(\mathfrak{g})\) and \(R_s(\mathfrak{g})\).  Assume that the joint unitary obeys
\begin{equation}
  [U_T,R_h(\mathfrak{g})\otimes R_s(\mathfrak{g})]=0
  \qquad \forall \mathfrak{g}\in \G .
  \label{S:eq:symmetry}
\end{equation}
Define
\begin{equation}
  \sigma_s^{(\mathfrak \mathfrak{g})}=R_s(\mathfrak{g})\sigma_s R_s^\dagger(\mathfrak{g}).
\end{equation}
For any harvester input \(\rho\),
\begin{align}
  \Phi_{\sigma_s^{(\mathfrak g)}}(\rho)
  &=
  \Tr_s\!\left[
    U_T(\rho\otimes R_s\sigma_s R_s^\dagger)U_T^\dagger
  \right]                                                   \\
  &=
  \Tr_s\!\left[
    U_T
    (R_h\otimes R_s)
    (R_h^\dagger\rho R_h\otimes\sigma_s)
    (R_h^\dagger\otimes R_s^\dagger)
    U_T^\dagger
  \right]                                                   \\
  &=
  \Tr_s\!\left[
    (R_h\otimes R_s)
    U_T
    (R_h^\dagger\rho R_h\otimes\sigma_s)
    U_T^\dagger
    (R_h^\dagger\otimes R_s^\dagger)
  \right]                                                   \\
  &=
  R_h(\mathfrak{g})
  \Tr_s\!\left[
    U_T
    (R_h^\dagger(\mathfrak{g})\rho R_h(\mathfrak{g})\otimes\sigma_s)
    U_T^\dagger
  \right]
  R_h^\dagger(\mathfrak{g})                                            \\
  &=
  R_h(\mathfrak{g})
  \Phi_{\sigma_s}\!\left[
    R_h^\dagger(\mathfrak{g})\rho R_h(\mathfrak{g})
  \right]
  R_h^\dagger(\mathfrak{g}).
  \label{S:eq:covariance}
\end{align}
In the fourth line we used invariance of the partial trace under unitary
conjugation on the traced subsystem.

For a connected Lie group, Eq.~\eqref{S:eq:symmetry} follows from additive
conserved charges
\begin{equation}
  Q^\alpha=Q_h^\alpha\otimes I_s+I_h\otimes Q_s^\alpha
\end{equation}
satisfying
\begin{equation}
  [H,Q^\alpha]=0.
\end{equation}
Then \(R_h(\mathfrak{g})\otimes R_s(\mathfrak{g})=\exp(i\sum_a\theta^a Q^\alpha)\) commutes with
\(U_T=e^{-iHT}\).

Now assume that \({\sigma_s}\in\Ddeh(\rho_0\to\rho_1)\), i.e.
\begin{equation}
  \Phi_{\sigma_s}(\rho_0)=\rho_1.
\end{equation}
Assume also that both boundary states are invariant under the harvester
representation,
\begin{equation}
  R_h(\mathfrak{g})\rho_0R_h^\dagger(\mathfrak{g})=\rho_0,\qquad
  R_h(\mathfrak{g})\rho_1R_h^\dagger(\mathfrak{g})=\rho_1 .
  \label{S:eq:boundary}
\end{equation}
Putting \(\rho=\rho_0\) in Eq.~\eqref{S:eq:covariance} gives
\begin{align}
  \Phi_{\sigma_s^{(\mathfrak g)}}(\rho_0)
  &=
  R_h(\mathfrak{g})
  \Phi_{\sigma_s}\!\left[
    R_h^\dagger(\mathfrak{g})\rho_0 R_h(\mathfrak{g})
  \right]
  R_h^\dagger(\mathfrak{g})                                         \\
  &=
  R_h(\mathfrak{g})\Phi_{\sigma_s}(\rho_0)R_h^\dagger(\mathfrak{g})                \\
  &=
  R_h(\mathfrak g)\rho_1R_h^\dagger(\mathfrak{g})                             \\
  &=
  \rho_1.
\end{align}
Therefore
\begin{equation}
  {\sigma_s}\in\Ddeh(\rho_0\to\rho_1)
  \quad\Rightarrow\quad
  \{R_s(\mathfrak{g}){\sigma_s} R_s^\dagger(\mathfrak{g}):\mathfrak{g}\in \G\}
  \subseteq
  \Ddeh(\rho_0\to\rho_1).
\end{equation}

The orbit is nontrivial when \({\sigma_s}\) is not invariant under the source
representation.  The stabilizer subgroup is
\begin{equation}
  \stab_\G({\sigma_s})
  =
  \{\mathfrak{g}\in \G:R_s(\mathfrak{g}){\sigma_s} R_s^\dagger(\mathfrak{g})={\sigma_s}\},
\end{equation}
and physically distinct orbit points are labeled by the relative complement \(\G \setminus \stab_\G({\sigma_s})\).

For an infinitesimal transformation generated by \(Q_s^\alpha\),
\begin{equation}
  {\sigma_s}(\theta)
  =
  e^{i\theta Q_s^\alpha}{\sigma_s} e^{-i\theta Q_s^\alpha},
\end{equation}
so
\begin{equation}
  \left.\frac{d{\sigma_s}(\theta)}{d\theta}\right|_{\theta=0}
  =
  i[Q_s^\alpha,{\sigma_s}].
\end{equation}
Thus the orbit is locally nontrivial whenever
\begin{equation}
  [Q_s^\alpha,{\sigma_s}]\neq0
\end{equation}
for at least one generator.

\section{Jaynes--Cummings realization of the Noether--DEH theorem}

We now show explicitly how the Noether--DEH theorem operates in the Jaynes--Cummings model. The source is a single quantized bosonic mode with annihilation operator \(a\), while the harvester is a two-level system with ground state \(\ket{g}\), excited state \(\ket{e}\), and raising and lowering operators
\begin{equation}
    \sigma_+ = \ket{e}\bra{g}, 
    \qquad 
    \sigma_- = \ket{g}\bra{e}.
\end{equation}
On resonance, and in the interaction picture with \(\hbar=1\), the Jaynes--Cummings Hamiltonian is
\begin{equation}
    H_{\rm JC}
    =
    g\left(a\sigma_+ + a^\dagger \sigma_-\right),
\end{equation}
where \(g\) is the light--matter coupling strength. This Hamiltonian exchanges one excitation between the field and the two-level harvester. Thus, the atom can be charged only by removing an excitation from the source, and the coherent oscillation between these two alternatives is the microscopic mechanism by which energy is deposited in the harvester.

The relevant Noether charge is the total excitation number
\begin{equation}
    N = n_s+n_h 
    =
    a^\dagger a + \ket{e}\bra{e}.
\end{equation}
A direct calculation gives
\begin{equation}
    [H_{\rm JC},N]=0.
\end{equation}
This conservation law is the Jaynes--Cummings version of the symmetry condition required by the Noether--DEH theorem. It implies invariance under the \(U(1)\) transformation
\begin{equation}
    R(\theta)=R_s(\theta)\otimes R_h(\theta),
    \qquad
    R_s(\theta)=e^{i\theta a^\dagger a},
    \qquad
    R_h(\theta)=e^{i\theta \ket{e}\bra{e}}.
\end{equation}
Physically, \(R_s(\theta)\) is a phase rotation of the field, while \(R_h(\theta)\) is the corresponding phase rotation of the excited-state amplitude of the harvester. Since the total excitation number is conserved, applying this phase rotation before the interaction is equivalent to applying it after the interaction. This is precisely the covariance structure underlying the Noether--DEH theorem.

Let
\begin{equation}
    U(\tau)=e^{-iH_{\rm JC}\tau}
\end{equation}
be the evolution at the harvesting time \(\tau\), and define the reduced harvester channel associated with a source state \(\sigma\) by
\begin{equation}
    \Phi_\sigma(\rho_h)
    =
    \Tr_s\!\left[
    U(\tau)(\rho_h\otimes \sigma)U(\tau)^\dagger
    \right].
\end{equation}
For the phase-rotated source state
\begin{equation}
    \sigma_\theta
    =
    R_s(\theta)\sigma R_s^\dagger(\theta)
    =
    e^{i\theta a^\dagger a}\sigma e^{-i\theta a^\dagger a},
\end{equation}
the conserved excitation number gives the covariance identity
\begin{equation}
    \Phi_{\sigma_\theta}(\rho_h)
    =
    R_h(\theta)\,
    \Phi_\sigma\!\left[
    R_h^\dagger(\theta)\rho_h R_h(\theta)
    \right]
    R_h^\dagger(\theta).
\end{equation}
This equation has a simple physical meaning. A phase change of the source does not produce a new, unrelated harvester dynamics. Rather, it produces the same dynamics viewed in the correspondingly rotated harvester phase frame.

For deterministic charging we take the harvester boundary states to be
\begin{equation}
    \rho_0=\ket{g}\bra{g},
    \qquad
    \rho_1=\ket{e}\bra{e}.
\end{equation}
Both are invariant under the local harvester symmetry,
\begin{equation}
    R_h(\theta)\rho_0R_h^\dagger(\theta)=\rho_0,
    \qquad
    R_h(\theta)\rho_1R_h^\dagger(\theta)=\rho_1.
\end{equation}
Although the vector \(\ket{e}\) acquires a phase under \(R_h(\theta)\), the density operator \(\ket{e}\bra{e}\) does not. Therefore, the deterministic charging task is insensitive to the conjugate phase generated by \(n_h\). If a source state \(\sigma\) satisfies
\begin{equation}
    \Phi_\sigma(\ket{g}\bra{g})
    =
    \ket{e}\bra{e},
\end{equation}
then covariance immediately gives
\begin{align}
    \Phi_{\sigma_\theta}(\ket{g}\bra{g})
    &=
    R_h(\theta)\Phi_\sigma(\ket{g}\bra{g})R_h^\dagger(\theta)
    \nonumber\\
    &=
    R_h(\theta)\ket{e}\bra{e}R_h^\dagger(\theta)
    \nonumber\\
    &=
    \ket{e}\bra{e}.
\end{align}
Thus every phase-rotated source \(\sigma_\theta\) performs the same deterministic transition. The deposited work,
\begin{equation}
    W_h
    =
    \Tr\!\left[
    H_h
    \left(
    \ket{e}\bra{e}-\ket{g}\bra{g}
    \right)
    \right],
\end{equation}
is consequently identical for all members of the orbit. The source phase may change the field state and the final source state, but it cannot appear as randomness in the final state of the harvester.

It is useful to see this statement at the level of the Jaynes--Cummings Rabi solution. For each photon number \(n\geq 1\), the Hamiltonian couples the two states \(\ket{g,n}\) and \(\ket{e,n-1}\), and
\begin{equation}
    U(t)\ket{g,n}
    =
    \cos(g\sqrt{n}\,t)\ket{g,n}
    -
    i\sin(g\sqrt{n}\,t)\ket{e,n-1}.
\end{equation}
A number state \(\ket{n}\) therefore perfectly excites the harvester at times satisfying
\begin{equation}
    g\sqrt{n}\,T
    =
    \frac{(2k_n+1)\pi}{2},
    \qquad
    k_n\in\mathbb{Z}_{\geq 0}.
\end{equation}
More generally, if a set of photon numbers \(\mathcal{M}_T\) is synchronized at the same harvesting time \(\tau\), meaning that every \(n\in\mathcal{M}_T\) satisfies the condition above, then any normalized source state
\begin{equation}
    \ket{\psi_s}
    =
    \sum_{n\in\mathcal{M}_T}c_n\ket{n}
\end{equation}
is a coherent DEH seed. Indeed,
\begin{equation}
    U(\tau)\left(\ket{g}\otimes\ket{\psi_s}\right)
    =
    \ket{e}\otimes
    \sum_{n\in\mathcal{M}_T}
    \left[-i(-1)^{k_n}c_n\right]\ket{n-1}.
\end{equation}
The harvester factorizes in the excited state, while all dependence on the amplitudes \(c_n\) is transferred to the final source state. Hence the field may contain coherence between different photon-number sectors, yet the harvester still ends in the pure state \(\ket{e}\bra{e}\).

The Noether--DEH theorem now generates a continuous family from this single seed. The phase-rotated seed is
\begin{equation}
    \ket{\psi_s(\theta)}
    =
    e^{i\theta a^\dagger a}\ket{\psi_s}
    =
    \sum_{n\in\mathcal{M}_T}c_n e^{in\theta}\ket{n}.
\end{equation}
At the same harvesting time,
\begin{equation}
    U(\tau)\left(\ket{g}\otimes\ket{\psi_s(\theta)}\right)
    =
    \ket{e}\otimes
    \sum_{n\in\mathcal{M}_\tau}
    \left[-i(-1)^{k_n}c_n e^{in\theta}\right]\ket{n-1}.
\end{equation}
Thus the phase \(\theta\) changes the relative phases in the source and in the final field state, but it leaves the harvester output exactly unchanged. The orbit is physically nontrivial whenever the seed has coherence between distinct photon-number sectors, equivalently whenever
\begin{equation}
    \left[
    a^\dagger a,
    \ket{\psi_s}\bra{\psi_s}
    \right]
    \neq 0.
\end{equation}
If the seed is photon-number diagonal, the phase orbit is trivial; if it contains photon-number coherence, the same conservation law generates a continuum of distinct DEH-capable source states.

As a concrete example, choose
\begin{equation}
    \tau=\frac{\pi}{2g}.
\end{equation}
Then the synchronized photon numbers include
\begin{equation}
    n=1,9,25,\ldots,
\end{equation}
because \(\sqrt{n}\) is an odd integer. The state
\begin{equation}
    \ket{\psi_s}
    =
    \frac{\ket{1}+\ket{9}}{\sqrt{2}}
\end{equation}
is therefore a nontrivial coherent seed. Its phase orbit is
\begin{equation}
    \ket{\psi_s(\theta)}
    =
    \frac{
    e^{i\theta}\ket{1}
    +
    e^{i9\theta}\ket{9}
    }{\sqrt{2}},
\end{equation}
and at time \(\tau=\pi/(2g)\) one obtains
\begin{equation}
    U(\tau)\left(\ket{g}\otimes\ket{\psi_s(\theta)}\right)
    =
    \ket{e}\otimes
    \frac{
    -ie^{i\theta}\ket{0}
    +
    ie^{i9\theta}\ket{8}
    }{\sqrt{2}}.
\end{equation}
All values of \(\theta\) therefore produce the same deterministic harvester transition
\begin{equation}
    \ket{g}\bra{g}
    \longrightarrow
    \ket{e}\bra{e},
\end{equation}
even though the source states are distinct for different \(\theta\). The phase information is not destroyed; it remains in the final field state. What the theorem guarantees is that this information is invisible to the reduced harvester output.

The same conclusion holds for classical uncertainty over the source phase. For any probability density \(p(\theta)\), define
\begin{equation}
    \bar{\sigma}
    =
    \int d\theta\,p(\theta)
    \ket{\psi_s(\theta)}\bra{\psi_s(\theta)}.
\end{equation}
Linearity of the reduced channel gives
\begin{align}
    \Phi_{\bar{\sigma}}(\ket{g}\bra{g})
    &=
    \int d\theta\,p(\theta)
    \Phi_{\ket{\psi_s(\theta)}\bra{\psi_s(\theta)}}(\ket{g}\bra{g})
    \nonumber\\
    &=
    \ket{e}\bra{e}.
\end{align}
Therefore, even a source with an unknown phase drawn from the Noether orbit charges the harvester deterministically. The harvester receives a fixed quantum of energy, while carrying no information about which phase-rotated source was used.

This example also clarifies the role and limitation of the theorem. The Noether--DEH theorem does not by itself find the synchronized seed; that is a dynamical condition set by the Jaynes--Cummings Rabi frequencies \(g\sqrt{n}\). Once such a seed is found, however, the conserved excitation number promotes it to an entire symmetry orbit of DEH-capable sources. In the Jaynes--Cummings model, deterministic phase-insensitive harvesting is therefore not an accidental feature of a special solution. It is the operational consequence of excitation-number conservation together with the phase invariance of the harvester boundary states.

\section{Three-spin XX-chain seed}
The principle is constructive: find one coherent DEH seed, then act on it with
the conserved source charge.  Consider a three-spin XX chain \(L-q-R\), where
the central spin \(q\) is the harvester and the edge spins form the source. Spin chains are standard platforms for coherent excitation transfer and quantum-battery models~\cite{Christandl2004,Le2018}.

In the resonant interaction picture,
\begin{equation}
  H
  =
  J\!\left[
  \sigma_q^+(\sigma_L^-+\sigma_R^-)
  +
  \sigma_q^-(\sigma_L^++\sigma_R^+)
  \right].
  \label{eq:xx}
\end{equation}
This Hamiltonian conserves
\[
  N=n_q+n_L+n_R.
\]
This makes the full system evolution invariant under the symmetry transformation
\begin{equation} R(\theta)= R_{LR}(\theta) \otimes R_q(\theta)  =  e^{i \theta (n_L + n_R+ n_q)}.
\end{equation}
Using {either brute force numerics, informed guesses, or analytic computations in some cases, one can often find some source state that can achieve DEH. In this case for instance,}

\begin{equation}
  |\psi_s\rangle_{LR} =   \frac{1}{\sqrt{3}} \left( |10\rangle + |01 \rangle + |11\rangle \right),
\end{equation}
{is a DEH enabling source}. In other words, the initial state \[
|\psi\rangle = |0\rangle_q \otimes 
|\psi_s\rangle_{LR} \]
 evolves into a state of the form, 
 \begin{equation} \rho_\tau = |1\rangle \langle 1|_q \otimes \rho_{LR},
\end{equation}
at the time $\tau=\frac{\pi}{2\sqrt2J}$.

Consequently every transformed source state mixture
\begin{align}
  \rho_\theta & = \int p(\theta) R_s(\theta) |\psi_s \rangle \langle \psi_s |_{LR} R_s^\dagger(\theta) d \theta \\
  & = \int \frac{p(\theta)}{3} d\theta [ \left(|10\rangle + |01 \rangle + e^{i \theta}|11\rangle\right) \nonumber \\ 
  &\hspace{2.8cm} \left(\langle 10| + \langle 01 | + e^{-i \theta} \langle 11| \right) ]
\end{align}
is simultaneously DEH achieving for any $p(\theta)$.

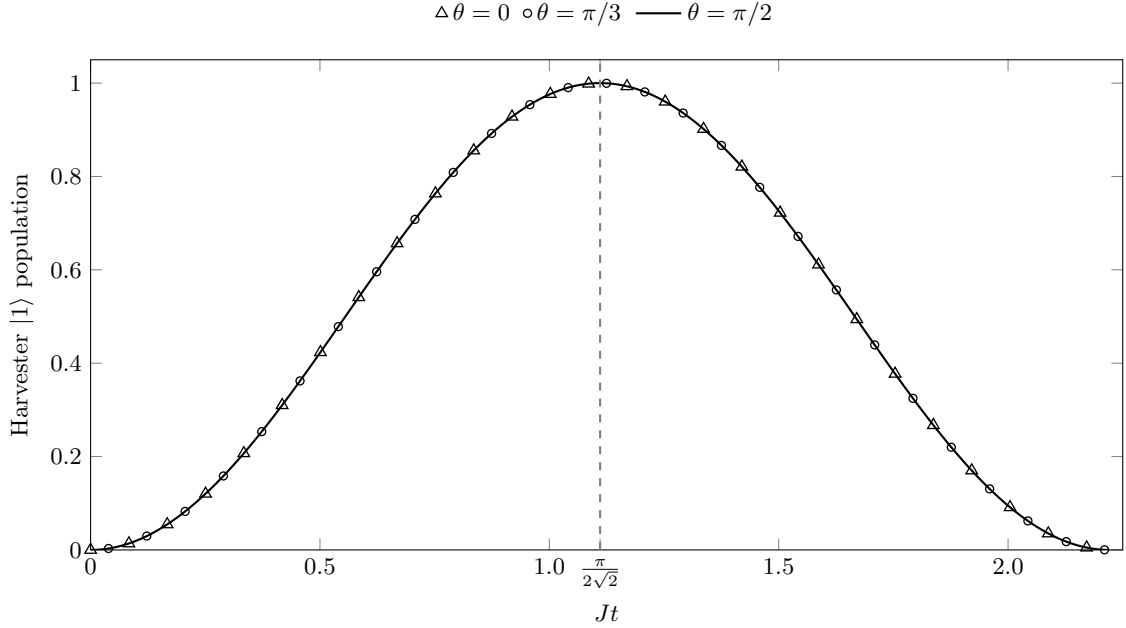
\begin{figure}[htbp!]
    \centering
        \begin{tikzpicture}
\begin{axis}[
    width=0.85\linewidth,
    height=0.45\linewidth,
    xlabel={\(J t\)},
    ylabel={Harvester \(|1\rangle\) population},
    xmin=0, xmax=2.25,
    ymin=0, ymax=1.05,
    xtick={0, 0.5, 1, 1.1107, 1.5, 2.0},
    xticklabels={0, 0.5, 1.0, \(\frac{\pi}{2\sqrt{2}}\), 1.5, 2.0}, 
    ytick distance=0.2,
    legend style={draw=none, at={(0.5,1.05)}, anchor=south, font=\small, legend columns=-1},
    tick label style={font=\small},
    label style={font=\small},
]

\addplot [
    only marks,
    mark=triangle*,
    mark size=2.5pt,
    draw=black,
    fill opacity=0,
    line width=0.5pt,
    mark repeat=15, 
    mark phase=1 
] 
table [x=t, y=fidelity1, col sep=space] {Three_spin_XX_chain_simulation_data.dat};
\addlegendentry{\(\theta=0\) \quad}

\addplot [
    only marks,
    mark=*,
    mark size=1.5pt,
    draw=black!60!black,
    fill opacity=0,
    line width=0.5pt,
    mark repeat=15,
    mark phase=8 
] 
table [x=t, y=fidelity2, col sep=space] {Three_spin_XX_chain_simulation_data.dat};
\addlegendentry{\(\theta=\pi/3\) \quad}

\addplot [
    black,
    thick
] 
table [x=t, y=fidelity3, col sep=space] {Three_spin_XX_chain_simulation_data.dat};
\addlegendentry{\(\theta=\pi/2\)}

\draw[dashed, black!50, thick] (axis cs:1.1107,0) -- (axis cs:1.1107,1.05);

\end{axis}
\end{tikzpicture}
        \caption{\justifying
        Three spin XX chain example.}
        \label{fig:DEH_spin_chain}
\end{figure}

Here, fig~\ref{fig:DEH_spin_chain} gives a visualization of the example. The source state is given by $|\psi_s\rangle_{LR}$, and under different rotation angles, the harvester has the same $|1\rangle$ popoulation.

This example also clarifies what is special about the seed.  The two source
number sectors {($n=1$ and $n=2$)} are dynamically synchronized: they have the same transfer time {$\tau= \frac{\pi}{2\sqrt 2 J}$} 
into the harvester.  Once such a synchronized pair is found, coherence between
the sectors supplies a symmetry-breaking seed {$R_s(\theta)\sigma_s R^\dagger_s(\theta) \neq \sigma_s $}, and Noether symmetry supplies
the continuous family.  The method is therefore not tied to the 
XX spin chain model; it is a recipe for Jaynes--Cummings, spin networks, multimode cavities, and other
symmetry-constrained quantum systems relevant to quantum-battery architectures~\cite{Ferraro2018,Le2018,CampaioliRMP2024}.

\section{Numerical details for the non-Abelian source-orbit example}

This section gives the reproducible numerical construction used in the $SU(2)$ example of the main text. The calculation has two parts. First, we find one high-fidelity source seed by singular-value optimization. Second, we verify that rotating this seed by $R_s(\mathcal R)=u(\mathcal R)^{\otimes5}$ leaves the pure-target harvester fidelity unchanged.

\begin{figure}[t]
 \resizebox{\linewidth}{!}{%
\begin{tikzpicture}[
    thick,
    font=\large, 
    spin/.style={circle, fill=black, inner sep=1.5pt},
    spin arrow/.style={->, >=Stealth, very thick}
]


\node at (0, 5.5) {harvester (h)};

\draw (-0.5, 4) -- (0.5, 4) node[right] {\(|e\rangle\)};

\draw (-2.2, 2.5) -- (-1.2, 2.5) node[above=4pt, pos=0.5] {\(|a, 1\rangle\)};
\draw (-0.5, 2.5) -- (0.5, 2.5) node[above=4pt, pos=0.5] {\(|a, 2\rangle\)};
\draw (1.2, 2.5) -- (2.2, 2.5) node[above=4pt, pos=0.5] {\(|a, 3\rangle\)};

\draw (-0.5, 1) -- (0.5, 1) node[right] {\(|g\rangle\)};

\draw[decorate, decoration={brace, amplitude=10pt, mirror}] (-2.5, 0.2) -- (2.5, 0.2);

\node[anchor=north] at (0, -0.3) {
    \(\begin{aligned}
        H_h &= \Omega_e |e\rangle\langle e| \\
            &\quad + \Omega_a \sum_\alpha |a, \alpha\rangle\langle a, \alpha| \\
            &\quad + 0 |g\rangle\langle g|
    \end{aligned}\)
};


\node at (6.5, 5.5) {source (s)};

\begin{scope}[shift={(3.5, 2.5)}]
    \node[spin] (s1) at (0,0) {};
    \draw[spin arrow] (-0.3, -0.5) -- (0.4, 0.6);
    
    \node[spin] (s2) at (1.5,0) {};
    \draw[spin arrow] (1.9, -0.5) -- (1.0, 0.6);
    
    \node[spin] (s3) at (3.0,0) {};
    \draw[spin arrow] (3.0, 0.6) -- (3.0, -0.7);
    
    \node[spin] (s4) at (4.5,0) {};
    \draw[spin arrow] (4.5, -0.6) -- (4.5, 0.7);
    
    \node[spin] (s5) at (6.0,0) {};
    \draw[spin arrow] (6.7, 0.3) -- (5.3, -0.3);
\end{scope}

\draw[decorate, decoration={brace, amplitude=10pt, mirror}] (3.0, 0.2) -- (10.0, 0.2);

\node[anchor=north] at (6.5, -0.7) {\(H_s = \sum_i J_i \vec{S}_i \cdot \vec{S}_{i+1}\)};

\end{tikzpicture}}  \caption{{\bf Non-abelian DEH example.} There is a 5 dimensional harvester and a $2^5$-dimensional source. }
  \label{fig:schematic}
\end{figure}
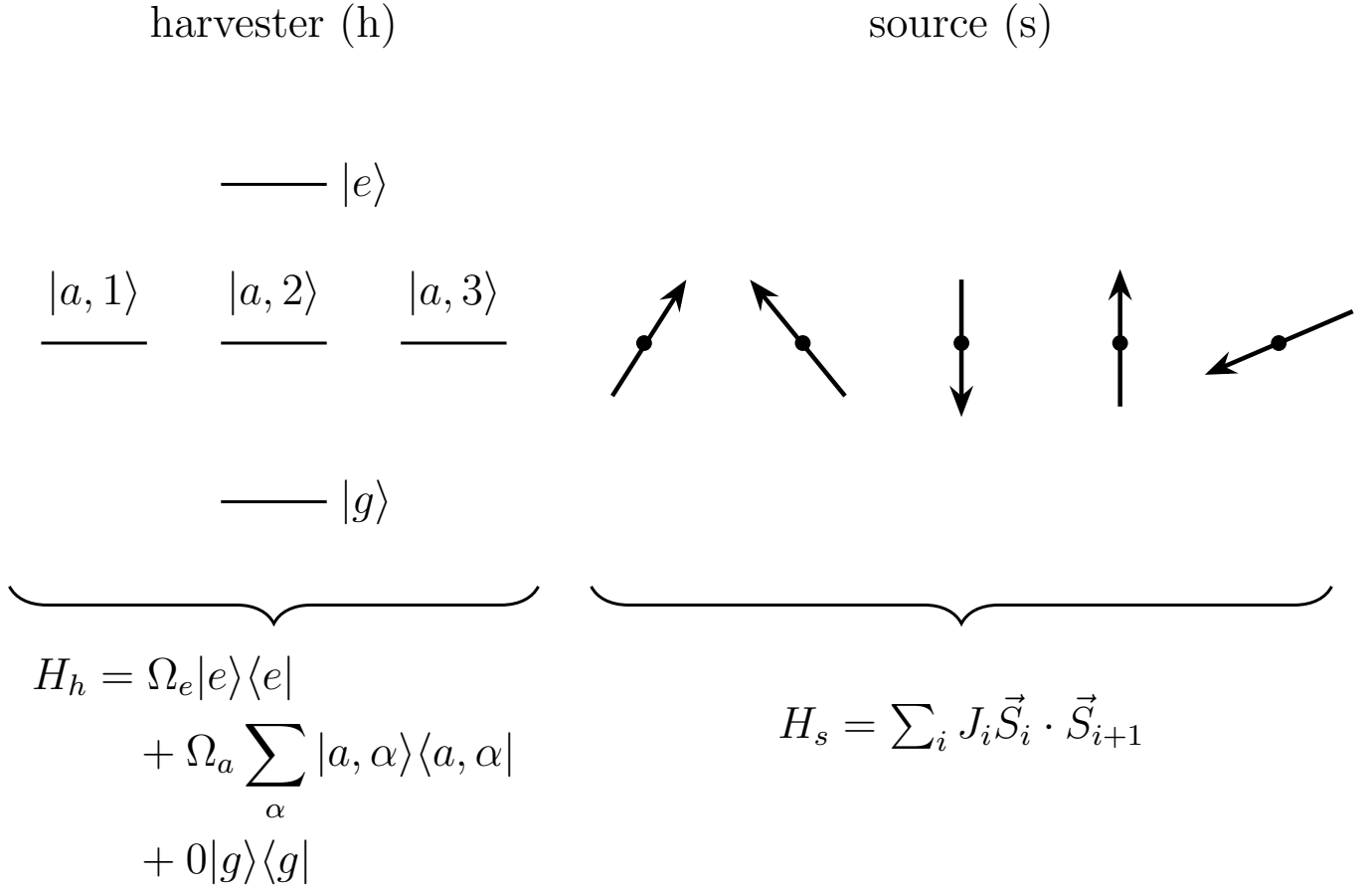

\subsection{Model}

The harvester basis is
\begin{equation}
\{|g\rangle,|a,x\rangle,|a,y\rangle,|a,z\rangle,|e\rangle\}.
\end{equation}
The states $|g\rangle$ and $|e\rangle$ are rotational scalars, while $|a,\alpha\rangle$ transform as a Cartesian spin-one triplet. The source contains five spin-$1/2$ sites. The Hamiltonian is
\begin{equation}
H=H_h\otimes I_s+I_h\otimes H_s+H_{\rm int},
\end{equation}
with
\begin{equation}
H_h=\Omega_a \Pi_a+\Omega_e|e\rangle\langle e|,
\qquad
\Pi_a=\sum_{\alpha=x,y,z}|a,\alpha\rangle\langle a,\alpha|,
\end{equation}
\begin{equation}
H_s=\sum_{i=1}^{4}J_i{\bf S}_i\cdot{\bf S}_{i+1},
\end{equation}
and
\begin{equation}
H_{\rm int}
=
\lambda_g\sum_{\alpha=x,y,z}D_g^\alpha\otimes T_s^\alpha
+
\lambda_e\sum_{\alpha=x,y,z}D_e^\alpha\otimes W_s^\alpha .
\end{equation}
Here
\begin{equation}
D_g^\alpha=|a,\alpha\rangle\langle g|+|g\rangle\langle a,\alpha|,
\qquad
D_e^\alpha=|a,\alpha\rangle\langle e|+|e\rangle\langle a,\alpha|,
\end{equation}
and
\begin{equation}
{\bf T}_s=\sum_{i=1}^{5}c_i{\bf S}_i,
\qquad
{\bf W}_s=\sum_{i=1}^{5}d_i{\bf S}_i .
\end{equation}

The harvester spin generators vanish on $|g\rangle$ and $|e\rangle$. On the auxiliary triplet they are
\begin{equation}
(J_h^\gamma)_{\alpha\beta}=-i\epsilon_{\gamma\alpha\beta},
\qquad
\alpha,\beta,\gamma\in\{x,y,z\}.
\end{equation}
The source generators are
\begin{equation}
J_s^\alpha=\sum_{i=1}^{5}S_i^\alpha .
\end{equation}
The numerical implementation verifies $
\frac{\|[H,J_h^\alpha+J_s^\alpha]\|}{\|H\|}$,
for $\alpha=x,y,z$.

\subsection{Seed search}

The transfer operator from the harvester ground state to the charged state is
\begin{equation}
A_{eg}(T)={}_h\langle e|e^{-iHT}|g\rangle_h .
\end{equation}
It acts only on the source Hilbert space. The pure-target fidelity is
\begin{equation}
F_e(T,\psi)=
\|A_{eg}(T)|\psi\rangle_s\|^2 .
\end{equation}
Hence
\begin{equation}
F_e^{\rm opt}(T)=s_{\max}^2[A_{eg}(T)].
\end{equation}
The seed is chosen as the dominant right singular vector of $A_{eg}(\tau_\star)$, where $\tau_\star$ maximizes $s_{\max}[A_{eg}(T)]$.

For the representative instance used in the main text, we take $J_1=1$
as the energy unit and choose
\begin{align}
(J_1,J_2,J_3,J_4)
&=
(1,\;0.003,\;-0.0024,\;2),
\\
(c_1,c_2,c_3,c_4,c_5)
&=
(1,\;-1,\;0.0024,\;-0.0018,\;0.0015),
\\
(d_1,d_2,d_3,d_4,d_5)
&=
(0.0012,\;-0.0015,\;1,\;0.0018,\;-0.0021),
\\
\Omega_a&=0.9985121238,
\\
\Omega_e&=0.9992837576,
\\
\lambda_g&=3.173487813,
\\
\lambda_e&=3.664428086.
\end{align}
Here $\lambda_e/\lambda_g=2/\sqrt{3}$. All energies and
frequencies are expressed in units of $J_1$, and $\hbar=1$.

For these parameters, direct numerical evaluation gives
\begin{equation}
\frac{\|[H,J_h^x+J_s^x]\|_\infty}{\|H\|_\infty}
=
1.36\times10^{-16},
\end{equation}
\begin{equation}
\frac{\|[H,J_h^y+J_s^y]\|_\infty}{\|H\|_\infty}
=
1.37\times10^{-16},
\end{equation}
and
\begin{equation}
\frac{\|[H,J_h^z+J_s^z]\|_\infty}{\|H\|_\infty}
=
1.77\times10^{-16},
\end{equation}
confirming the $SU(2)$ symmetry to numerical precision.

The optimal transfer time obtained by maximizing
$s_{\max}[A_{eg}(T)]$ is
\begin{equation}
\tau_\star=0.7000333028 .
\end{equation}
Equivalently,
\begin{equation}
\Omega_e\tau_\star=0.6995319093.
\end{equation}
At this time,
\begin{equation}
s_{\max}[A_{eg}(\tau_\star)]
=
0.9999896432,
\end{equation}
and hence
\begin{equation}
F_e(\tau_\star,\psi_\star)
=
s_{\max}^2[A_{eg}(\tau_\star)]
=
0.9999792865.
\label{eq:su2_numerical_fidelity}
\end{equation}
The corresponding infidelity is therefore
\begin{equation}
1-F_e(\tau_\star,\psi_\star)
=
2.07135\times10^{-5}.
\end{equation}

The four largest singular values are degenerate to numerical precision,
\begin{equation}
s_1=s_2=s_3=s_4=0.9999896432,
\end{equation}
and the corresponding right-singular subspace forms a single
spin-$3/2$ multiplet. We choose $|\psi_\star\rangle_s$ to be the
highest-weight state within this dominant singular subspace. It then
satisfies
\begin{equation}
{\bf J}_s^2|\psi_\star\rangle_s
=
\frac{15}{4}|\psi_\star\rangle_s,
\qquad
J_s^z|\psi_\star\rangle_s
=
\frac{3}{2}|\psi_\star\rangle_s .
\label{eq:su2_seed_spin}
\end{equation}
Since $J=3/2>0$, the seed is not invariant under the full $SU(2)$
action and therefore has a nontrivial source orbit.

For completeness, in the computational basis
$|b_1b_2b_3b_4b_5\rangle_s$, with $|0\rangle$ and $|1\rangle$
denoting the $S^z=+1/2$ and $S^z=-1/2$ states, respectively, the
chosen highest-weight seed is
\begin{align}
|\psi_\star\rangle_s
={}&
(-0.000832522+0.001254597\,i)|00001\rangle
\nonumber\\
&+
(0.000587308-0.001745229\,i)|00010\rangle
\nonumber\\
&+
0.816575268\,|00100\rangle
\nonumber\\
&+
(-0.407499275+0.000873728\,i)|01000\rangle
\nonumber\\
&+
(-0.408830778-0.000383096\,i)|10000\rangle .
\label{eq:su2_seed_explicit}
\end{align}
The overall phase has been chosen such that the coefficient of
$|00100\rangle$ is real and positive.

At the harvesting time, the populations of the harvester ground,
auxiliary, and charged sectors are
\begin{align}
P_g(\tau_\star)&=1.11770\times10^{-5},
\\
P_a(\tau_\star)&=9.53646\times10^{-6},
\\
P_e(\tau_\star)&=0.9999792865,
\end{align}
where
\begin{equation}
P_a=\sum_{\alpha=x,y,z}
\langle a,\alpha|\rho_h|a,\alpha\rangle .
\end{equation}
Thus the residual population in the auxiliary harvester triplet is
below $10^{-5}$ for this approximate seed. During the transfer the
auxiliary sector is appreciably occupied: its maximum population is
approximately
\begin{equation}
P_a^{\rm max}=0.500052
\end{equation}
at
\begin{equation}
t\simeq0.350017,
\end{equation}
showing that the auxiliary levels participate in the dynamics but are
almost completely depopulated again at the harvesting time.

\subsection{Rotated source states}

The source rotation is
\begin{equation}
R_s(\alpha,\beta,\gamma)
=
e^{-i\alpha J_s^z}
e^{-i\beta J_s^y}
e^{-i\gamma J_s^z}.
\end{equation}
The rotated source family is
\begin{equation}
|\psi_\star(\alpha,\beta,\gamma)\rangle_s
=
R_s(\alpha,\beta,\gamma)|\psi_\star\rangle_s .
\end{equation}
In the total-spin basis,
\begin{equation}
|\psi_\star(\alpha,\beta,\gamma)\rangle_s
=
\sum_{J,\mu,m,m'}
C_{J\mu m}
D^{(J)}_{m'm}(\alpha,\beta,\gamma)
|J,\mu,m'\rangle_s .
\end{equation}
Thus the orbit is nontrivial whenever the seed has support in $J>0$ sectors.

In the computational basis,
\begin{equation}
|\psi_\star\rangle_s=\sum_{\bf b}C_{\bf b}|{\bf b}\rangle_s,
\end{equation}
and
\begin{equation}
|\psi_\star(\alpha,\beta,\gamma)\rangle_s
=
u(\alpha,\beta,\gamma)^{\otimes5}
\sum_{\bf b}C_{\bf b}|{\bf b}\rangle_s .
\end{equation}
A generic rotation therefore changes the full pattern of five-spin amplitudes.

For the rotations used in the main-text figure we obtain
\begin{equation}
\begin{array}{c|c|c}
(\alpha,\beta,\gamma)
&
F_e(\tau_\star,\psi_\star(\alpha,\beta,\gamma))
&
|\langle\psi_\star|
\psi_\star(\alpha,\beta,\gamma)\rangle|^2
\\ \hline
(0,0,0)
&
0.999979286535
&
1
\\
(0.2,0.7,1.1)
&
0.999979286535
&
0.687112174
\\
(1.0,0.4,2.2)
&
0.999979286535
&
0.886203529
\\
(2.0,1.3,0.5)
&
0.999979286535
&
0.254538050
\end{array}
\label{eq:su2_rotated_fidelities}
\end{equation}
The harvesting fidelities agree to better than $10^{-12}$, whereas
the overlaps between the source states can differ substantially. In
particular, for the last rotation the squared overlap with the
unrotated seed is only approximately $0.255$, while the harvester
dynamics is unchanged. This directly illustrates that the symmetry
orbit contains physically distinct many-body source preparations
that implement the same harvesting task.

\subsection{Code structure}

The calculation can be implemented with the following steps.

\begin{enumerate}
\item Build the five-spin source operators $S_i^x,S_i^y,S_i^z$ and the total spin operators $J_s^x,J_s^y,J_s^z$.
\item Build $H_s$, ${\bf T}_s$, and ${\bf W}_s$.
\item Build the five-dimensional harvester Hilbert space and the operators $H_h$, $D_g^\alpha$, $D_e^\alpha$, and $J_h^\alpha$.
\item Construct the full Hamiltonian $H$ and verify Hermiticity.
\item Verify the three symmetry commutators $[H,J_h^\alpha+J_s^\alpha]$.
\item Diagonalize $H$ once.
\item Use the eigendecomposition to construct $A_{eg}(T)$ efficiently for many times.
\item Maximize $s_{\max}[A_{eg}(T)]$ over $T$.
\item Extract the dominant right singular vector $|\psi_\star\rangle_s$.
\item Apply several source rotations and verify that the fidelity is unchanged.
\end{enumerate}

A compact implementation is:

\begin{verbatim}
# Build H and check [H,J_h^a+J_s^a]=0.

evals, evecs = scipy.linalg.eigh(H)

# Harvester basis: |g>, |a,x>, |a,y>, |a,z>, |e>
cols_g = source_indices_in_harvester_block(g_index)
rows_e = source_indices_in_harvester_block(e_index)

Vrow = evecs[rows_e, :]
Vcol = evecs.conj().T[:, cols_g]

def Aeg(T):
    phases = np.exp(-1j*evals*T)
    return (Vrow*phases) @ Vcol

def smax(T):
    return scipy.linalg.svdvals(Aeg(T))[0]

# Search for Tstar.
Tstar = maximize_smax_over_time()

Astar = Aeg(Tstar)
U_svd, svals, Vh = scipy.linalg.svd(Astar, full_matrices=False)

# Dominant four-dimensional singular subspace (J=3/2 multiplet)
Vtop = Vh.conj().T[:, :4]

# Choose the Jz = +3/2 highest-weight state inside that subspace
Jz_restricted = Vtop.conj().T @ Jsz @ Vtop
mvals, Q = scipy.linalg.eigh(Jz_restricted)
psi_seed = Vtop @ Q[:, np.argmax(mvals)]

# Fix an irrelevant global phase for reproducible printed amplitudes
kmax = np.argmax(np.abs(psi_seed))
psi_seed *= np.exp(-1j * np.angle(psi_seed[kmax]))

# Fidelity of this particular seed
F_seed = np.vdot(
    psi_seed,
    Astar.conj().T @ Astar @ psi_seed
).real

# Rotate source.
def R_source(alpha,beta,gamma):
    return expm(-1j*alpha*Jsz) @ expm(-1j*beta*Jsy) @ expm(-1j*gamma*Jsz)

M = Astar.conj().T @ Astar
for angles in angles_list:
    psi_rot = R_source(*angles) @ psi_seed
    F_rot = np.vdot(psi_rot, M @ psi_rot).real
    print(angles, F_rot)
\end{verbatim}

In an exact DEH instance, $F_e=1$ implies that the final reduced harvester state is the pure state $|e\rangle\langle e|$. Hence the auxiliary triplet carries no residual population or entropy at the harvesting time. In an approximate numerical instance, $1-F_e$ is the leakage out of the desired pure final state and should be quoted explicitly.

\section{Approximate DEH States under Exact Symmetry}
\label{app:approx_states}

In the main text, the Noether-DEH theorem assumes that the seed source state $\sigma_s$ perfectly implements the desired DEH transition, i.e.,
\begin{equation}
    \Phi_{\sigma_s}(\rho_h^i)=\rho_h^f.
\end{equation}
Here, we consider the robust behavior of the protocol when the underlying dynamical symmetry and boundary-state invariances are exact, but the seed source state $\sigma_s$ is only an \textit{approximate} DEH state.

Let the joint system possess an exact symmetry such that the induced harvester channel satisfies the exact covariance identity
\begin{equation}
    \Phi_{\sigma_s^{(\mathfrak g)}}(\rho_h)
    =
    R_h(\mathfrak g)
    \Phi_{\sigma_s}
    \left[
        R_h^\dagger(\mathfrak g)\rho_h R_h(\mathfrak g)
    \right]
    R_h^\dagger(\mathfrak g),
    \label{eq:exact_cov}
\end{equation}
where $\sigma_s^{(\mathfrak g)}$ is the transformed source state defined in the main text. Furthermore, we assume that the harvester boundary states $\rho_h^i$ and $\rho_h^f$ are exactly invariant under the harvester symmetry action:
\begin{equation}
    R_h(\mathfrak g)\rho_h^iR_h^\dagger(\mathfrak g)=\rho_h^i,
    \qquad
    R_h(\mathfrak g)\rho_h^fR_h^\dagger(\mathfrak g)=\rho_h^f.
    \label{eq:exact_boundary}
\end{equation}

Suppose that the seed source state $\sigma_s$ achieves the target transition up to a trace-norm error bounded by $\epsilon_{\rm seed}\geq 0$:
\begin{equation}
    \left\|
        \Phi_{\sigma_s}(\rho_h^i)-\rho_h^f
    \right\|_1
    \leq
    \epsilon_{\rm seed}.
\end{equation}
We now evaluate the performance of any transformed source state $\sigma_s^{(\mathfrak g)}$ along the symmetry orbit. Using Eq.~\eqref{eq:exact_cov} and the invariance of $\rho_h^i$ from Eq.~\eqref{eq:exact_boundary}, its harvesting error is
\begin{align}
    \left\|
        \Phi_{\sigma_s^{(\mathfrak g)}}(\rho_h^i)-\rho_h^f
    \right\|_1
    &=
    \left\|
        R_h(\mathfrak g)
        \Phi_{\sigma_s}
        \left[
            R_h^\dagger(\mathfrak g)\rho_h^iR_h(\mathfrak g)
        \right]
        R_h^\dagger(\mathfrak g)
        -
        \rho_h^f
    \right\|_1
    \nonumber\\
    &=
    \left\|
        R_h(\mathfrak g)
        \Phi_{\sigma_s}(\rho_h^i)
        R_h^\dagger(\mathfrak g)
        -
        \rho_h^f
    \right\|_1.
\end{align}
Using the exact invariance of the target state,
\begin{equation}
    \rho_h^f
    =
    R_h(\mathfrak g)\rho_h^fR_h^\dagger(\mathfrak g),
\end{equation}
we can rewrite this expression as
\begin{align}
    \left\|
        \Phi_{\sigma_s^{(\mathfrak g)}}(\rho_h^i)-\rho_h^f
    \right\|_1
    &=
    \left\|
        R_h(\mathfrak g)
        \Phi_{\sigma_s}(\rho_h^i)
        R_h^\dagger(\mathfrak g)
        -
        R_h(\mathfrak g)\rho_h^fR_h^\dagger(\mathfrak g)
    \right\|_1
    \nonumber\\
    &=
    \left\|
        R_h(\mathfrak g)
        \left[
            \Phi_{\sigma_s}(\rho_h^i)-\rho_h^f
        \right]
        R_h^\dagger(\mathfrak g)
    \right\|_1.
\end{align}
Because the trace norm is unitarily invariant,
\begin{equation}
    \left\|UAU^\dagger\right\|_1=\|A\|_1,
\end{equation}
we obtain
\begin{equation}
    \left\|
        \Phi_{\sigma_s^{(\mathfrak g)}}(\rho_h^i)-\rho_h^f
    \right\|_1
    =
    \left\|
        \Phi_{\sigma_s}(\rho_h^i)-\rho_h^f
    \right\|_1
    \leq
    \epsilon_{\rm seed}.
\end{equation}

Thus, under exact dynamical symmetry and exact boundary-state invariance, the harvesting error is constant along the entire source symmetry orbit. If the seed source state approximately implements the desired transition within a tolerance $\epsilon_{\rm seed}$, every transformed source state $\sigma_s^{(\mathfrak g)}$ implements the transition with exactly the same accuracy. The symmetry therefore protects the protocol from any performance degradation along the orbit.

\section{Approximate symmetry}
\label{sec:approx_symmetry_derivation}

The exact theorem assumes exact covariance. A simple robustness statement
follows when covariance is approximate. Suppose that for all \(\mathfrak g\)
in a set of interest,
\begin{equation}
    \left\|
        \Phi_{\sigma_s^{(\mathfrak g)}}(\rho_h^i)
        -
        R_h(\mathfrak g)
        \Phi_{\sigma_s}
        \left(
            R_h^\dagger(\mathfrak g)
            \rho_h^i
            R_h(\mathfrak g)
        \right)
        R_h^\dagger(\mathfrak g)
    \right\|_1
    \leq
    \epsilon_{\rm cov},
\end{equation}
and that the boundary states are approximately invariant,
\begin{equation}
    \left\|
        R_h^\dagger(\mathfrak g)
        \rho_h^i
        R_h(\mathfrak g)
        -
        \rho_h^i
    \right\|_1
    \leq
    \epsilon_i,
    \qquad
    \left\|
        R_h(\mathfrak g)
        \rho_h^f
        R_h^\dagger(\mathfrak g)
        -
        \rho_h^f
    \right\|_1
    \leq
    \epsilon_f.
\end{equation}
If
\begin{equation}
    \Phi_{\sigma_s}(\rho_h^i)=\rho_h^f,
\end{equation}
then contractivity of quantum channels gives
\begin{equation}
    \left\|
        \Phi_{\sigma_s^{(\mathfrak g)}}(\rho_h^i)
        -
        \rho_h^f
    \right\|_1
    \leq
    \epsilon_{\rm cov}+\epsilon_i+\epsilon_f.
\end{equation}
Thus weak symmetry breaking or weak boundary non-invariance produces a
controlled deviation from exact deterministic harvesting.

Let the continuous symmetry group representation acting on the joint system
be given by the unitary operator
\begin{equation}
    R(\mathfrak g)
    =
    R_h(\mathfrak g)\otimes R_s(\mathfrak g)
    =
    e^{i\theta Q},
\end{equation}
where
\begin{equation}
    Q=Q_h+Q_s
\end{equation}
is the total charge generator. Suppose the symmetry is weakly broken by a
perturbation operator \(V\), with bounded spectral norm
\(\|V\|_\infty\leq 1\), such that the total Hamiltonian \(H\) satisfies
\begin{equation}
    [H,Q]=\epsilon V,
\end{equation}
where \(\epsilon\ll 1\) represents the symmetry-breaking scale.

To determine how this affects the joint time-evolution operator
\(U(\tau)=e^{-iH\tau}\), we evaluate the rotated Hamiltonian using the
Hadamard lemma:
\begin{equation}
    R^\dagger(\mathfrak g)HR(\mathfrak g)
    =
    e^{-i\theta Q}He^{i\theta Q}
    =
    H-i\theta[Q,H]
    +
    \mathcal{O}(\epsilon\theta^2).
\end{equation}
Substituting \([Q,H]=-\epsilon V\), the transformed Hamiltonian becomes a
perturbed operator:
\begin{equation}
    R^\dagger(\mathfrak g)HR(\mathfrak g)
    =
    H+i\theta\epsilon V
    +
    \mathcal{O}(\epsilon\theta^2).
\end{equation}
Using the property of the matrix exponential, the rotated time-evolution
operator can be expressed directly in terms of this perturbed generator:
\begin{equation}
    R^\dagger(\mathfrak g)U(\tau) R(\mathfrak g)
    =
    e^{-i\left(R^\dagger(\mathfrak g)HR(\mathfrak g)\right)\tau}
    =
    e^{-i(H+i\theta\epsilon V)\tau}
    +
    \mathcal{O}(\epsilon\theta^2).
\end{equation}
Using Duhamel's formula (equivalently, the first-order Dyson expansion), the
time evolution under this time-independent perturbed generator can be
expanded linearly in the perturbation \(i\theta\epsilon V\):
\begin{equation}
    e^{-i(H+i\theta\epsilon V)\tau}
    =
    U(\tau)
    +
    \theta\epsilon
    \int_0^\tau dt\,
    e^{-iH(\tau-t)}Ve^{-iHt}
    +
    \mathcal{O}(\epsilon\theta^2).
\end{equation}
Notice that since \(\mathcal{O}(\epsilon^2\theta^2)\) is subleading
comparing to \(\mathcal{O}(\epsilon\theta^2)\), it is omitted in the
perturbation expansion. Now, multiplying both sides by \(R(\mathfrak g)\)
from the left yields the fundamental tracking of the broken covariance at
the unitary level:
\begin{equation}
    U(\tau) R(\mathfrak g)
    -
    R(\mathfrak g)U(\tau)
    =
    \delta U,
\end{equation}
where the non-commutation unitary error operator \(\delta U\) is explicitly
defined to leading order as
\begin{equation}
    \delta U
    \approx
    \theta\epsilon R(\mathfrak g)
    \int_0^\tau dt\,
    e^{-iH(\tau-t)}Ve^{-iHt}.
\end{equation}
Taking the spectral/operator norm \(\|\cdot\|_\infty\) and applying the
triangle inequality, we find that the unitary tracking error scales
linearly with the rotation angle \(\theta\) and the harvesting time
\(\tau\):
\begin{equation}
    \|\delta U\|_\infty
    \leq
    |\theta|\epsilon
    \int_0^\tau dt\,\|V\|_\infty
    =
    |\theta|\tau\epsilon.
\end{equation}

We now evaluate the covariance discrepancy operator \(\mathcal{D}\) for the
induced harvester channel acting on the initial state \(\rho_h^i\),
defined as
\begin{equation}
    \mathcal{D}
    =
    \Phi_{\sigma_s^{(\mathfrak g)}}(\rho_h^i)
    -
    R_h(\mathfrak g)
    \Phi_{\sigma_s}
    \left(
        R_h^\dagger(\mathfrak g)
        \rho_h^i
        R_h(\mathfrak g)
    \right)
    R_h^\dagger(\mathfrak g),
\end{equation}
where
\begin{equation}
    \sigma_s^{(\mathfrak g)}
    =
    R_s(\mathfrak g)
    \sigma_s
    R_s^\dagger(\mathfrak g)
\end{equation}
is the symmetry-transformed source state.

Using the identity that the partial trace over the source system eliminates
any local source rotations,
\begin{equation}
    \Tr_s
    \left[
        R_s(\mathfrak g)X R_s^\dagger(\mathfrak g)
    \right]
    =
    \Tr_s[X],
\end{equation}
the second term can be rewritten globally as
\begin{align}
    &R_h(\mathfrak g)
    \Phi_{\sigma_s}
    \left(
        R_h^\dagger(\mathfrak g)
        \rho_h^i
        R_h(\mathfrak g)
    \right)
    R_h^\dagger(\mathfrak g)
    \nonumber\\
    &\quad=
    \Tr_s
    \left[
        R(\mathfrak g)U(\tau)
        \left(
            R_h^\dagger(\mathfrak g)
            \rho_h^i
            R_h(\mathfrak g)
            \otimes
            \sigma_s
        \right)
        U(\tau)^\dagger
        R^\dagger(\mathfrak g)
    \right]
    \nonumber\\
    &\quad=
    \Tr_s
    \left[
        \left(
            R(\mathfrak g)U(\tau) R^\dagger(\mathfrak g)
        \right)
        \Omega
        \left(
            R(\mathfrak g)U(\tau) R^\dagger(\mathfrak g)
        \right)^\dagger
    \right],
\end{align}
where
\begin{equation}
    \Omega
    =
    \rho_h^i\otimes\sigma_s^{(\mathfrak g)}
\end{equation}
is the joint unrotated harvester and rotated source state. Substituting
\begin{equation}
    R(\mathfrak g)U(\tau) R^\dagger(\mathfrak g)
    =
    U(\tau)-\delta U R^\dagger(\mathfrak g)
\end{equation}
into the partial trace yields
\begin{align}
    \mathcal{D}
    =
    \Tr_s\Big[
        U(\tau)\Omega U(\tau)^\dagger
        &-
        \left(
            U(\tau)-\delta U R^\dagger(\mathfrak g)
        \right)
        \Omega
        \nonumber\\
        &\times
        \left(
            U(\tau)-\delta U R^\dagger(\mathfrak g)
        \ \right)^\dagger
    \Big].
\end{align}
Expanding this expression and keeping only first-order terms in
\(\epsilon\) results in
\begin{equation}
    \mathcal{D}
    \approx
    \Tr_s
    \left[
        \delta U R^\dagger(\mathfrak g)
        \Omega U(\tau)^\dagger
        +
        U(\tau)\Omega R(\mathfrak g)\delta U^\dagger
    \right].
\end{equation}
We take the trace norm \(\|\cdot\|_1\) of the discrepancy. Because the
partial trace is non-increasing under the trace norm,
\(\|\Tr_s[X]\|_1\leq\|X\|_1\), we apply the triangle inequality and
H\"older's inequality:
\begin{align}
    \|\mathcal{D}\|_1
    &\leq
    \left\|
        \delta U R^\dagger(\mathfrak g)
        \Omega U(\tau)^\dagger
    \right\|_1
    +
    \left\|
        U(\tau)\Omega R(\mathfrak g)\delta U^\dagger
    \right\|_1
    \nonumber\\
    &\leq
    2
    \|\delta U\|_\infty
    \|R(\mathfrak g)\|_\infty
    \|U(\tau)\|_\infty
    \|\Omega\|_1.
\end{align}
Noting that \(R(\mathfrak g)\) and \(U(\tau)\) are unitary,
\begin{equation}
    \|R(\mathfrak g)\|_\infty
    =
    \|U(\tau)\|_\infty
    =
    1,
\end{equation}
and that the state \(\Omega\) is normalized,
\(\|\Omega\|_1=1\), substituting the bound for
\(\|\delta U\|_\infty\) directly reveals the microscopic scaling of the
channel covariance error:
\begin{equation}
    \|\mathcal{D}\|_1
    \leq
    2|\theta|\tau\epsilon
    \equiv
    \epsilon_{\rm cov}.
\end{equation}

Finally, we incorporate the potential non-invariance of the boundary states.
Suppose that the initial state \(\rho_h^i\) and target state
\(\rho_h^f\) are only approximately invariant under the symmetry group
action, such that
\begin{equation}
    \left\|
        R_h^\dagger(\mathfrak g)
        \rho_h^i
        R_h(\mathfrak g)
        -
        \rho_h^i
    \right\|_1
    \leq
    \epsilon_i,
    \qquad
    \left\|
        R_h(\mathfrak g)
        \rho_h^f
        R_h^\dagger(\mathfrak g)
        -
        \rho_h^f
    \right\|_1
    \leq
    \epsilon_f.
\end{equation}
Furthermore, we do not assume that the unrotated seed source state
\(\sigma_s\) achieves perfect deterministic harvesting; instead, it deviates
from the target state by a small baseline error parameter
\(\epsilon_{\rm seed}\):
\begin{equation}
    \left\|
        \Phi_{\sigma_s}(\rho_h^i)
        -
        \rho_h^f
    \right\|_1
    \leq
    \epsilon_{\rm seed}.
\end{equation}
We evaluate the operational failure of the transformed source state
\(\sigma_s^{(\mathfrak g)}\) to reach the target state \(\rho_h^f\) via
the triangle inequality:
\begin{align}
    \left\|
        \Phi_{\sigma_s^{(\mathfrak g)}}(\rho_h^i)
        -
        \rho_h^f
    \right\|_1
    &\leq
    \left\|
        \Phi_{\sigma_s^{(\mathfrak g)}}(\rho_h^i)
        -
        R_h(\mathfrak g)
        \Phi_{\sigma_s}
        \left(
            R_h^\dagger(\mathfrak g)
            \rho_h^i
            R_h(\mathfrak g)
        \right)
        R_h^\dagger(\mathfrak g)
    \right\|_1
    \nonumber\\
    &\quad+
    \left\|
        R_h(\mathfrak g)
        \Phi_{\sigma_s}
        \left(
            R_h^\dagger(\mathfrak g)
            \rho_h^i
            R_h(\mathfrak g)
        \right)
        R_h^\dagger(\mathfrak g)
        -
        \rho_h^f
    \right\|_1.
\end{align}
The first term is bounded exactly by the derived channel covariance bound
\(\epsilon_{\rm cov}\). For the second term, we use the unitary invariance
of the trace norm to factor out \(R_h(\mathfrak g)\) on the left and
\(R_h^\dagger(\mathfrak g)\) on the right:
\begin{align}
    &\left\|
        R_h(\mathfrak g)
        \Phi_{\sigma_s}
        \left(
            R_h^\dagger(\mathfrak g)
            \rho_h^i
            R_h(\mathfrak g)
        \right)
        R_h^\dagger(\mathfrak g)
        -
        \rho_h^f
    \right\|_1
    \nonumber\\
    &\quad=
    \left\|
        \Phi_{\sigma_s}
        \left(
            R_h^\dagger(\mathfrak g)
            \rho_h^i
            R_h(\mathfrak g)
        \right)
        -
        R_h^\dagger(\mathfrak g)
        \rho_h^f
        R_h(\mathfrak g)
    \right\|_1.
\end{align}
To evaluate this expression without assuming exact deterministic energy
harvesting, we insert both \(\Phi_{\sigma_s}(\rho_h^i)\) and
\(\rho_h^f\) as intermediate states. Applying the triangle inequality:
\begin{align}
    &\left\|
        \Phi_{\sigma_s}
        \left(
            R_h^\dagger(\mathfrak g)
            \rho_h^i
            R_h(\mathfrak g)
        \right)
        -
        R_h^\dagger(\mathfrak g)
        \rho_h^f
        R_h(\mathfrak g)
    \right\|_1
    \nonumber\\
    &\quad\leq
    \left\|
        \Phi_{\sigma_s}
        \left(
            R_h^\dagger(\mathfrak g)
            \rho_h^i
            R_h(\mathfrak g)
        \right)
        -
        \Phi_{\sigma_s}(\rho_h^i)
    \right\|_1
    \nonumber\\
    &\qquad+
    \left\|
        \Phi_{\sigma_s}(\rho_h^i)
        -
        \rho_h^f
    \right\|_1
    \nonumber\\
    &\qquad+
    \left\|
        \rho_h^f
        -
        R_h^\dagger(\mathfrak g)
        \rho_h^f
        R_h(\mathfrak g)
    \right\|_1.
\end{align}
By the contractivity of quantum channels under the trace norm,
\begin{equation}
    \|\Phi_{\sigma_s}(A)-\Phi_{\sigma_s}(B)\|_1
    \leq
    \|A-B\|_1,
\end{equation}
the first term simplifies to the initial-state non-invariance bound
\(\epsilon_i\). The second term is explicitly bounded by our baseline seed
harvesting error \(\epsilon_{\rm seed}\). The third term corresponds
directly to the final target-state non-invariance bound \(\epsilon_f\).
Therefore, we arrive at the generalized, robust operational error bound:
\begin{equation}
    \left\|
        \Phi_{\sigma_s^{(\mathfrak g)}}(\rho_h^i)
        -
        \rho_h^f
    \right\|_1
    \leq
    \epsilon_{\rm seed}
    +
    \epsilon_{\rm cov}
    +
    \epsilon_i
    +
    \epsilon_f
    \leq
    2|\theta|\tau\epsilon
    +
    \epsilon_{\rm seed}
    +
    \epsilon_i
    +
    \epsilon_f.
\end{equation}

\section{Generalised phase-space formulation}
\label{sec:generalised-phase-space}

The symmetry argument developed above is not tied to the Hilbert-space
representation of quantum mechanics. It can be formulated directly in
phase space and therefore applies to classical Hamiltonian mechanics, the
Wigner--Weyl formulation of quantum mechanics, and more general
phase-space theories.

Let the joint phase space of the harvester \(h\) and source \(s\) be
\begin{equation}
    \Gamma=\Gamma_h\times\Gamma_s,
\end{equation}
with phase-space coordinates \(z_h\in\Gamma_h\) and
\(z_s\in\Gamma_s\). A state of the composite system is represented by a
normalised real function
\begin{equation}
    f(z_h,z_s,t).
\end{equation}
Depending on the theory, \(f\) may be a classical probability density, a
Wigner quasiprobability distribution, or a state in a more general
phase-space representation.

We consider dynamics generated by the generalised bracket
\begin{equation}
    \frac{\partial f}{\partial t}
    =
    \{f,H\}_{K}
    \equiv
    \mathcal L_H f,
    \label{eq:generalised-eom}
\end{equation}
where \(H(z_h,z_s)\) is the joint Hamiltonian and
\begin{equation}
    \{A,B\}_{K}
    :=
    \int d\kappa\,K(\kappa)\,
    A\sin\!\left(\frac{\kappa}{2}\overleftrightarrow{\Lambda}\right)B.
    \label{eq:generalised-bracket}
\end{equation}
Here \(K(\kappa)\) specifies the dynamical theory, while
\begin{equation}
    \overleftrightarrow{\Lambda}=\overleftrightarrow{\Lambda}_h+\overleftrightarrow{\Lambda}_s
\end{equation}
is the bidifferential symplectic operator on the composite phase space.
For one canonical degree of freedom in each subsystem,
\begin{equation}
    \overleftrightarrow{\Lambda}_\alpha
    =
    \overleftarrow{\partial}_{x_\alpha}
    \overrightarrow{\partial}_{p_\alpha}
    -
    \overleftarrow{\partial}_{p_\alpha}
    \overrightarrow{\partial}_{x_\alpha},
    \qquad
    \alpha\in\{h,s\},
\end{equation}
with the natural extension to several degrees of freedom. Appropriate
choices of \(K\) recover, in particular, classical Poisson evolution and
quantum Moyal evolution.

For a time-independent Hamiltonian, let
\begin{equation}
    \mathcal T_T
    :=
    e^{T\mathcal L_H}
\end{equation}
denote the joint dynamical transformation at time \(T\), so that
\begin{equation}
    f(T)=\mathcal T_T f(0).
\end{equation}
The covariance argument below depends only on the resulting transformation
\(\mathcal T_T\), rather than on the explicit form of the bracket in
Eq.~\eqref{eq:generalised-bracket}. The bracket becomes relevant when the
dynamics and their continuous symmetries are expressed in infinitesimal
form.

Suppose that the source and harvester are initially uncorrelated:
\begin{equation}
    f(0)=f_{h,i}f_s,
\end{equation}
where \(f_{h,i}\) and \(f_s\) are the initial harvester and source states,
respectively.

Define the source-marginalisation map \(\mathsf M_s\) by
\begin{equation}
    \bigl(\mathsf M_s F\bigr)(z_h)
    :=
    \int_{\Gamma_s}dz_s\,F(z_h,z_s)
\end{equation}
for any integrable joint phase-space function \(F\). For a fixed source
state \(f_s\), the reduced harvester map at time \(T\) is
\begin{equation}
    \Phi_{f_s}^{T}[f_h]
    :=
    \mathsf M_s\mathcal T_T[f_hf_s]
    =
    \int_{\Gamma_s}dz_s\,
    \mathcal T_T[f_hf_s].
    \label{eq:reduced-phase-space-map}
\end{equation}

The source state \(f_s\) achieves deterministic energy harvesting for the
transition
\begin{equation}
    f_{h,i}\longrightarrow f_{h,f}
\end{equation}
at time \(T\) if
\begin{equation}
    \Phi_{f_s}^{T}[f_{h,i}]
    =
    f_{h,f}.
    \label{eq:phase-space-deh}
\end{equation}
The corresponding energy deposited in the harvester is
\begin{equation}
    W_h
    =
    \int_{\Gamma_h}dz_h\,H_h(z_h)
    \bigl[
        f_{h,f}(z_h)-f_{h,i}(z_h)
    \bigr],
    \label{eq:phase-space-work}
\end{equation}
where \(H_h\) is the free harvester Hamiltonian. Thus, as in the operator
formulation, DEH is a state-to-state task: the prescribed final harvester
state, and consequently the deposited energy, must be obtained at the
fixed harvesting time.

\subsection{Symmetry covariance of the reduced dynamics}

Let a group \(G\) act on the harvester and source state spaces through
reversible transformations
\begin{equation}
    \mathcal R_h(g),
    \qquad
    \mathcal R_s(g),
    \qquad
    g\in G.
\end{equation}
We denote the corresponding transformation of the composite state space by
\begin{equation}
    \mathcal R_{hs}(g)
    :=
    \mathcal R_h(g)\boxtimes\mathcal R_s(g).
    \label{eq:joint-gpt-action}
\end{equation}
Here \(\boxtimes\) denotes the product transformation defined by the
chosen composite theory; it need not be interpreted as a Hilbert-space
tensor product. On uncorrelated states, it satisfies
\begin{equation}
    \mathcal R_{hs}(g)[f_hf_s]
    =
    \bigl(\mathcal R_h(g)f_h\bigr)
    \bigl(\mathcal R_s(g)f_s\bigr).
    \label{eq:product-action}
\end{equation}

We also assume that the joint symmetry action is compatible with
marginalisation:
\begin{equation}
    \mathsf M_s\mathcal R_{hs}(g)
    =
    \mathcal R_h(g)\mathsf M_s.
    \label{eq:marginalisation-covariance}
\end{equation}
Explicitly, for every integrable joint function \(F(z_h,z_s)\),
\begin{equation}
    \int_{\Gamma_s}dz_s\,
    \bigl[\mathcal R_{hs}(g)F\bigr](z_h,z_s)
    =
    \mathcal R_h(g)
    \left[
        \int_{\Gamma_s}dz_s\,F(z_h,z_s)
    \right].
    \label{eq:measure-preservation}
\end{equation}
For transformations induced by canonical changes of phase-space
coordinates, this relation follows from invariance of the phase-space
integration measure.

The group \(G\) is a symmetry of the joint dynamics when
\begin{equation}
    \mathcal T_T\circ\mathcal R_{hs}(g)
    =
    \mathcal R_{hs}(g)\circ\mathcal T_T,
    \qquad
    \forall g\in G.
    \label{eq:general-phase-space-symmetry}
\end{equation}
This equation states that applying the symmetry transformation before the
dynamics gives the same result as applying it after the dynamics. It is an
equality between transformations under composition and does not assume a
Hilbert-space or unitary structure.

For a transformed source state
\begin{equation}
    f_{s,g}
    :=
    \mathcal R_s(g)f_s,
    \label{eq:transformed-source-phase-space}
\end{equation}
Eqs.~\eqref{eq:product-action}--\eqref{eq:general-phase-space-symmetry}
imply the covariance relation
\begin{equation}
    \Phi_{f_{s,g}}^{T}[f_h]
    =
    \mathcal R_h(g)
    \Phi_{f_s}^{T}
    \left[
        \mathcal R_h(g)^{-1}f_h
    \right].
    \label{eq:phase-space-covariance}
\end{equation}

Indeed, using the product structure of the symmetry action,
\begin{equation}
    f_h\,\mathcal R_s(g)f_s
    =
    \mathcal R_{hs}(g)
    \left[
        \bigl(\mathcal R_h(g)^{-1}f_h\bigr)f_s
    \right].
\end{equation}
It follows that
\begin{align}
    \Phi_{f_{s,g}}^{T}[f_h]
    &=
    \mathsf M_s\mathcal T_T
    \mathcal R_{hs}(g)
    \left[
        \bigl(\mathcal R_h(g)^{-1}f_h\bigr)f_s
    \right]
    \notag\\
    &=
    \mathsf M_s\mathcal R_{hs}(g)
    \mathcal T_T
    \left[
        \bigl(\mathcal R_h(g)^{-1}f_h\bigr)f_s
    \right]
    \notag\\
    &=
    \mathcal R_h(g)\mathsf M_s
    \mathcal T_T
    \left[
        \bigl(\mathcal R_h(g)^{-1}f_h\bigr)f_s
    \right]
    \notag\\
    &=
    \mathcal R_h(g)
    \Phi_{f_s}^{T}
    \left[
        \mathcal R_h(g)^{-1}f_h
    \right].
\end{align}
In the second line we used symmetry of the joint dynamics, while in the
third line we used compatibility of the symmetry with source
marginalisation.

\begin{theorem}[Generalised phase-space Noether--DEH theorem]
\label{thm:generalised-phase-space-deh}
Let \(\mathcal T_T\) be the joint phase-space evolution of a source and
harvester, and suppose that
\begin{equation}
    \mathcal T_T\circ\mathcal R_{hs}(g)
    =
    \mathcal R_{hs}(g)\circ\mathcal T_T,
    \qquad
    \forall g\in G.
\end{equation}
Assume that a source state \(f_s\) achieves the deterministic transition
\begin{equation}
    \Phi_{f_s}^{T}[f_{h,i}]
    =
    f_{h,f},
\end{equation}
and that the harvester boundary states are invariant under the local
symmetry action:
\begin{equation}
    \mathcal R_h(g)f_{h,i}
    =
    f_{h,i},
    \qquad
    \mathcal R_h(g)f_{h,f}
    =
    f_{h,f}.
    \label{eq:phase-space-boundary-invariance}
\end{equation}
Then every transformed source state
\begin{equation}
    f_{s,g}
    =
    \mathcal R_s(g)f_s
\end{equation}
achieves the same deterministic transition:
\begin{equation}
    \Phi_{f_{s,g}}^{T}[f_{h,i}]
    =
    f_{h,f}.
\end{equation}
\end{theorem}

\begin{proof}
Using the covariance relation in Eq.~\eqref{eq:phase-space-covariance}
and invariance of the initial harvester state,
\begin{align}
    \Phi_{f_{s,g}}^{T}[f_{h,i}]
    &=
    \mathcal R_h(g)
    \Phi_{f_s}^{T}
    \left[
        \mathcal R_h(g)^{-1}f_{h,i}
    \right]
    \notag\\
    &=
    \mathcal R_h(g)
    \Phi_{f_s}^{T}[f_{h,i}]
    \notag\\
    &=
    \mathcal R_h(g)f_{h,f}.
\end{align}
Invariance of the final harvester state then gives
\begin{equation}
    \Phi_{f_{s,g}}^{T}[f_{h,i}]
    =
    f_{h,f}.
\end{equation}
\end{proof}

Consequently, the complete source orbit
\begin{equation}
    \mathcal O_{f_s}^{G}
    :=
    \left\{
        \mathcal R_s(g)f_s
        \,\middle|\,
        g\in G
    \right\}
    \label{eq:phase-space-source-orbit}
\end{equation}
belongs to the set of source states implementing the same deterministic
harvesting task. The orbit is physically nontrivial whenever \(f_s\) is
not invariant under the source-side action.

\subsection{Continuous symmetries and generalised Noether charges}

For a one-parameter continuous group, let the symmetry be generated by an
additive phase-space observable
\begin{equation}
    Q(z_h,z_s)
    =
    Q_h(z_h)+Q_s(z_s).
\end{equation}
Define the corresponding generators acting on phase-space states by
\begin{equation}
    \mathcal L_Q f
    :=
    \{f,Q\}_{K},
    \qquad
    \mathcal L_{Q_h}f_h
    :=
    \{f_h,Q_h\}_{K},
    \qquad
    \mathcal L_{Q_s}f_s
    :=
    \{f_s,Q_s\}_{K}.
    \label{eq:symmetry-generators}
\end{equation}
The finite transformations are formally written as
\begin{equation}
    \mathcal R_{hs}(\theta)
    =
    e^{\theta\mathcal L_Q},
    \qquad
    \mathcal R_\alpha(\theta)
    =
    e^{\theta\mathcal L_{Q_\alpha}},
    \qquad
    \alpha\in\{h,s\}.
    \label{eq:finite-generalised-transformations}
\end{equation}

The infinitesimal symmetry condition corresponding to
Eq.~\eqref{eq:general-phase-space-symmetry} is
\begin{equation}
    [\mathcal L_H,\mathcal L_Q]
    =
    0.
    \label{eq:liouvillian-symmetry}
\end{equation}
It ensures that the symmetry flow commutes with the physical time
evolution:
\begin{equation}
    e^{T\mathcal L_H}e^{\theta\mathcal L_Q}
    =
    e^{\theta\mathcal L_Q}e^{T\mathcal L_H}.
\end{equation}

The invariance conditions for the harvester boundary states become
\begin{equation}
    \mathcal L_{Q_h}f_{h,i}
    =
    0,
    \qquad
    \mathcal L_{Q_h}f_{h,f}
    =
    0,
\end{equation}
or, equivalently,
\begin{equation}
    \{f_{h,i},Q_h\}_{K}
    =
    0,
    \qquad
    \{f_{h,f},Q_h\}_{K}
    =
    0.
\end{equation}
By contrast, the source orbit is nontrivial whenever
\begin{equation}
    \mathcal L_{Q_s}f_s
    =
    \{f_s,Q_s\}_{K}
    \neq
    0.
    \label{eq:nontrivial-source-orbit}
\end{equation}
Thus, a single DEH-achieving source state that breaks the source-side
symmetry generates a continuous family of distinct phase-space states
producing the same deterministic harvester output.

Care is required when relating the dynamical symmetry condition
\eqref{eq:liouvillian-symmetry} to conservation of the generalised charge.
Under the usual smoothness and boundary assumptions, the skew-adjointness
of the generalised bracket gives
\begin{align}
    \frac{d}{dt}\langle Q\rangle
    &=
    \int_{\Gamma}dz\,
    \{f,H\}_{K}\,Q
    \notag\\
    &=
    \int_{\Gamma}dz\,
    f\,\{H,Q\}_{K}.
    \label{eq:generalised-charge-conservation}
\end{align}
Therefore,
\begin{equation}
    \{H,Q\}_{K}=0
\end{equation}
implies conservation of the expectation value of \(Q\).

For the ordinary Poisson and Moyal brackets, the Jacobi identity further
implies
\begin{equation}
    [\mathcal L_H,\mathcal L_Q]
    =
    \mathcal L_{\{Q,H\}_{K}},
    \label{eq:jacobi-liouvillian-relation}
\end{equation}
with the sign determined by the convention
\(\mathcal L_A f=\{f,A\}_{K}\). In these cases,
\begin{equation}
    \{H,Q\}_{K}=0
\end{equation}
is sufficient to establish symmetry of the full evolution.

A generalised bracket, however, need not satisfy the Jacobi identity.
Conservation of \(\langle Q\rangle\) then does not by itself establish
that the transformation generated by \(Q\) commutes with the dynamics.
The generalised phase-space DEH theorem therefore requires the dynamical
covariance condition
\begin{equation}
    [\mathcal L_H,\mathcal L_Q]=0
\end{equation}
directly. Whenever the bracket satisfies the relevant Jacobi identity,
this condition follows from the familiar Noether relation
\begin{equation}
    \{H,Q\}_{K}=0.
\end{equation}

The physical content is nevertheless unchanged across the different
phase-space theories. A symmetry of the joint dynamics identifies a
direction in source-state space that is invisible to a
symmetry-invariant harvesting task. Moving the source along this
direction changes its phase-space state and may change the full joint
trajectory, but it leaves the final reduced harvester state and the
deposited energy exactly unchanged. Noether symmetry therefore organises
DEH-achieving states into continuous phase-space orbits in classical,
quantum, and generalised Hamiltonian theories.

\section{Classical DEH example with Noether's theorem}
\label{app:classical_model}

In Appendix~F of Ref.~\cite{MengPRA2025}, the harvester is a uniformly
charged rotating sphere driven by a prescribed time-dependent magnetic
field.  The purpose of this section is to promote that field to a dynamical
source with its own phase-space variables.  We first construct the joint
Hamiltonian and recover the externally driven model in the negligible-
backreaction limit.  We then identify its continuous symmetry and exhibit an
explicit deterministic energy harvesting trajectory.

\subsection{Externally driven rotating sphere, source quadratures and the joint Hamiltonian}

Let the sphere have moment of inertia $\mathfrak I_h$ and angular momentum
$\mathbf L_h=(L^x_h,L^y_h,L^z_h)$.  Its magnetic moment is taken to be proportional
to its angular momentum,
\begin{equation}
    \boldsymbol{\mu}=\beta\mathbf L_h,
    \label{eq:classical_mu_beta_L}
\end{equation}
where $\beta$ is a constant fixed by the charge and mass distribution.  In a
magnetic field $\mathbf B(t)$, the Hamiltonian is
\begin{equation}
    H_h(t)=\frac{\mathbf L_h^2}{2\mathfrak I_h}
    -\boldsymbol{\mu}\cdot\mathbf B(t)
    =\frac{\mathbf L_h^2}{2\mathfrak I_h}-\beta\mathbf L_h\cdot\mathbf B(t).
    \label{eq:classical_external_hamiltonian}
\end{equation}

The angular-momentum components satisfy
\begin{equation}
    \{L^i_h,L^j_h\}_h=\epsilon_{ijk}L^k_h.
    \label{eq:classical_L_poisson_components}
\end{equation}
It follows that
\begin{align}
    \dot{\mathbf L}_h
    &=\left(\frac{\mathbf L_h}{\mathfrak I_h}-\beta\mathbf B(t)\right)
      \times\mathbf L_h \nonumber \\
    &=\frac{\mathbf L_h}{\mathfrak I_h}\times\mathbf L_h
      -\beta\mathbf B(t)\times\mathbf L_h  \nonumber \\
    &=\beta\mathbf L_h\times\mathbf B(t),
    \label{eq:classical_external_L_eom}
\end{align}
where $\mathbf L_h\times\mathbf L_h=0$ was used in the last line.

For concreteness, consider a circularly rotating transverse field
\begin{equation}
    \mathbf B_{\perp}(t)
    =B_1\left[
       \cos(\omega t+\phi_0)\,\hat{\mathbf x}
       +\sin(\omega t+\phi_0)\,\hat{\mathbf y}
      \right].
    \label{eq:classical_external_rotating_field}
\end{equation}
The corresponding time-dependent interaction potential is
\begin{equation}
    V(t)
    =-\beta B_1\left[
       L^x_h\cos(\omega t+\phi_0)
       +L^y_h\sin(\omega t+\phi_0)
      \right].
    \label{eq:classical_external_rotating_potential}
\end{equation}
In Eqs.~\eqref{eq:classical_external_rotating_field} and
\eqref{eq:classical_external_rotating_potential}, the field is prescribed by
hand: it has no dynamical variables of its own and cannot respond to the
harvester.  We next replace it by an autonomous Hamiltonian source.

The simplest autonomous system producing sinusoidal motion is a harmonic
oscillator.  Let $X$ and $P$ denote scaled canonical quadratures of the
source, satisfying
\begin{equation}
    \{X,P\}_s=1.
    \label{eq:classical_XP_bracket}
\end{equation}
For example, starting from physical position and momentum variables
$X_{\rm phys}$ and $P_{\rm phys}$ for a source of mass $m_s$, one may define
\begin{equation}
    X=\sqrt{m_s\omega}\,X_{\rm phys},
    \qquad
    P=\frac{P_{\rm phys}}{\sqrt{m_s\omega}}.
    \label{eq:classical_scaled_quadratures}
\end{equation}
The free source Hamiltonian then takes the symmetric form
\begin{equation}
    H_s=\frac{\omega}{2}\left(X^2+P^2\right)
       \equiv\omega J_s,
    \qquad
    J_s:=\frac{X^2+P^2}{2},
    \label{eq:classical_source_hamiltonian}
\end{equation}
where $J_s$ is the source action.  Hamilton's equations for the uncoupled
source are
\begin{equation}
    \dot X=\frac{\partial H_s}{\partial P}=\omega P,
    \qquad
    \dot P=-\frac{\partial H_s}{\partial X}=-\omega X.
    \label{eq:classical_free_source_eom}
\end{equation}
Consequently, $\ddot X=-\omega^2X$.  For the initial condition
\begin{equation}
    X(0)=A\cos\phi_0,
    \qquad
    P(0)=-A\sin\phi_0,
    \label{eq:classical_source_initial_XP}
\end{equation}
the solution is
\begin{equation}
    X(t)=A\cos(\omega t+\phi_0),
    \qquad
    P(t)=-A\sin(\omega t+\phi_0).
    \label{eq:classical_source_free_solution}
\end{equation}
Thus the phase-space vector
\begin{equation}
    \mathbf b_s(X,P):=X\hat{\mathbf x}-P\hat{\mathbf y}
    \label{eq:classical_b_source_definition}
\end{equation}
obeys
\begin{equation}
    \mathbf b_s(t)
    =A\left[
       \cos(\omega t+\phi_0)\,\hat{\mathbf x}
       +\sin(\omega t+\phi_0)\,\hat{\mathbf y}
      \right].
    \label{eq:classical_b_source_free_solution}
\end{equation}
The pair $(X,-P)$ therefore supplies the cosine and sine quadratures of a
rotating field.  The sign multiplying $P$ is conventional: it follows from
the initial condition in Eq.~\eqref{eq:classical_source_initial_XP} and could
be reversed by changing the orientation assigned to the source phase.

We convert the source vector $\mathbf b_s$ into a magnetic field by defining
\begin{equation}
    \mathbf B_s(X,P)
    :=\frac{g}{\beta}\mathbf b_s(X,P)
    =\frac{g}{\beta}
      \left(X\hat{\mathbf x}-P\hat{\mathbf y}\right),
    \label{eq:classical_source_field_definition}
\end{equation}
where $g$ is a coupling constant that converts the chosen quadrature units
into magnetic-field units.  The interaction Hamiltonian is then
\begin{align}
    H_{\rm int}
    &:=-\boldsymbol{\mu}\cdot\mathbf B_s(X,P) \nonumber \\
    &=-\beta\mathbf L_h\cdot
      \frac{g}{\beta}
      \left(X\hat{\mathbf x}-P\hat{\mathbf y}\right) \nonumber \\
    &=-g\left(L^x_hX-L^y_hP\right).
    \label{eq:classical_interaction_hamiltonian}
\end{align}
In particular,
\begin{equation}
    L^x_hX-L^y_hP
    =\mathbf L_h\cdot(X,-P,0),
    \label{eq:classical_interaction_is_dot}
\end{equation}
so Eq.~\eqref{eq:classical_interaction_hamiltonian} is still the standard
magnetic-dipole dot-product interaction.

Along the free source orbit in Eq.~\eqref{eq:classical_source_free_solution},
Eq.~\eqref{eq:classical_source_field_definition} gives
\begin{equation}
    \mathbf B_s(t)
    =\frac{gA}{\beta}\left[
       \cos(\omega t+\phi_0)\,\hat{\mathbf x}
       +\sin(\omega t+\phi_0)\,\hat{\mathbf y}
      \right].
    \label{eq:classical_source_field_free_orbit}
\end{equation}
Comparison with Eq.~\eqref{eq:classical_external_rotating_field} shows that
\begin{equation}
    B_1=\frac{gA}{\beta},
    \qquad\text{or equivalently}\qquad
    g=\frac{\beta B_1}{A}.
    \label{eq:classical_g_identification}
\end{equation}
Thus $g$ does not represent an additional field beyond the one appearing in
the externally driven model.  It separates the phase-space amplitude $A$ of
the source from the conversion between source quadratures and magnetic-field
strength.  Alternatively, if $X$ and $-P$ are themselves expressed in
magnetic-field units, one may set $g=\beta$.

For the DEH task considered below, we allow the harvester to possess a storage term $\Omega L^z_h$.  The full autonomous
Hamiltonian is
\begin{equation}
    H=H_h+H_s+H_{\rm int},
    \label{eq:classical_total_decomposition}
\end{equation}
where
\begin{equation}
    H_h=\frac{\mathbf L_h^2}{2\mathfrak I_h}+\Omega L^z_h,
    \qquad
    H_s=\frac{\omega}{2}(X^2+P^2),
    \qquad
    H_{\rm int}=-g(L^x_hX-L^y_hP).
    \label{eq:classical_joint_hamiltonian_terms}
\end{equation}
Equivalently,
\begin{equation}
    \boxed{
    H=\frac{\mathbf L_h^2}{2\mathfrak I_h}
      +\Omega L^z_h
      +\frac{\omega}{2}(X^2+P^2)
      -g(L^x_hX-L^y_hP).}
    \label{eq:classical_joint_hamiltonian}
\end{equation}
If the storage term is generated by a static magnetic field
$B_0\hat{\mathbf z}$, then $\Omega=-\beta B_0$.  Setting $\Omega=0$ recovers
the unbiased rotating-sphere model.  In that unbiased case, a reversal of
$\mathbf L_h$ changes its orientation but not its free kinetic energy
$\mathbf L_h^2/(2\mathfrak I_h)$.  A nonzero $\Omega$, or another orientation-dependent
storage term, is therefore required if the transition is to represent a nonzero increase of harvester energy.

\subsection{Coupled equations of motion and recovery of the external model}
\label{app:classical_coupled_eom}

The construction above also clarifies the meaning of obtaining an effective
time-dependent potential from a joint Hamiltonian.  When the source follows a
specified trajectory $(X(t),P(t))$, the potential acting on the harvester is
the conditional interaction
\begin{equation}
    V_{X(t),P(t)}(\mathbf L_h)
    =H_{\rm int}(\mathbf L_h,X(t),P(t)).
    \label{eq:classical_conditional_potential}
\end{equation}
By contrast, derivatives such as $-\partial_XH_{\rm int}$ and
$-\partial_PH_{\rm int}$ determine the backreaction forces on the source.

For the harvester,
\begin{equation}
    \nabla_{\mathbf L_h}H
    =\frac{\mathbf L_h}{\mathfrak I_h}
     +\Omega\hat{\mathbf z}
     -g\left(X\hat{\mathbf x}-P\hat{\mathbf y}\right).
    \label{eq:classical_joint_gradient_L}
\end{equation}
Therefore,
\begin{align}
    \dot{\mathbf L}_h
    &=\nabla_{\mathbf L_h}H\times\mathbf L_h \nonumber \\
    &=\Omega\hat{\mathbf z}\times\mathbf L_h
      -g\left(X\hat{\mathbf x}-P\hat{\mathbf y}\right)
      \times\mathbf L_h \nonumber \\
    &=\Omega\hat{\mathbf z}\times\mathbf L_h
      +g\mathbf L_h\times
      \left(X\hat{\mathbf x}-P\hat{\mathbf y}\right).
    \label{eq:classical_joint_L_vector_eom}
\end{align}
The kinetic term drops out because
$(\mathbf L_h/\mathfrak I_h)\times\mathbf L_h=0$.  In components,
\begin{align}
    \dot L^x_h&=-\Omega L^y_h+gPL^z_h,
    \label{eq:classical_joint_Lx_eom}\\
    \dot L^y_h&=\Omega L^x_h+gXL^z_h,
    \label{eq:classical_joint_Ly_eom}\\
    \dot L^z_h&=-g(PL^x_h+XL^y_h).
    \label{eq:classical_joint_Lz_eom}
\end{align}

The source equations are
\begin{align}
    \dot X
    &=\frac{\partial H}{\partial P}
      =\omega P+gL^y_h,
    \label{eq:classical_joint_X_eom}\\
    \dot P
    &=-\frac{\partial H}{\partial X}
      =-\omega X+gL^x_h.
    \label{eq:classical_joint_P_eom}
\end{align}
The terms $gL^y_h$ and $gL^x_h$ are the backreaction of the harvester on the
source.  If these terms are negligible, Eqs.~\eqref{eq:classical_joint_X_eom}
and \eqref{eq:classical_joint_P_eom} reduce to the free solution in
Eq.~\eqref{eq:classical_source_free_solution}.  Substitution into
Eq.~\eqref{eq:classical_joint_L_vector_eom} then reproduces a sphere driven
by the prescribed rotating field in
Eq.~\eqref{eq:classical_external_rotating_field}.  When the backreaction is
retained, the same Hamiltonian instead describes a genuinely autonomous
source--harvester system.

The magnitude of the harvester angular momentum is conserved even in the
fully coupled model.  Indeed,
\begin{equation}
    \frac{d}{dt}\mathbf L_h^2
    =2\mathbf L_h\cdot\dot{\mathbf L}_h=0,
    \label{eq:classical_L_magnitude_conserved}
\end{equation}
because every term in Eq.~\eqref{eq:classical_joint_L_vector_eom} is
perpendicular to $\mathbf L_h$.  We therefore write
\begin{equation}
    |\mathbf L_h|=L_h
    \label{eq:classical_L_ell}
\end{equation}
throughout the remainder of the derivation.

\subsection{Continuous joint rotation and conserved charge}
\label{app:classical_noether_symmetry}

Consider a simultaneous rotation of the harvester and the source phase by an
angle $\theta$.  On the harvester variables, define
\begin{equation}
    \mathbf L_h\longmapsto R_z(\theta)\mathbf L_h,
    \label{eq:classical_harvester_rotation}
\end{equation}
where
\begin{equation}
    R_z(\theta)=
    \begin{pmatrix}
        \cos\theta&-\sin\theta&0\\
        \sin\theta& \cos\theta&0\\
        0&0&1
    \end{pmatrix}.
    \label{eq:classical_Rz_matrix}
\end{equation}
On the source quadratures, define
\begin{equation}
    \begin{pmatrix}X\\P\end{pmatrix}
    \longmapsto
    \begin{pmatrix}
       \cos\theta& \sin\theta\\
      -\sin\theta& \cos\theta
    \end{pmatrix}
    \begin{pmatrix}X\\P\end{pmatrix}.
    \label{eq:classical_source_rotation}
\end{equation}
Under Eq.~\eqref{eq:classical_source_rotation}, the vector
$\mathbf b_s=(X,-P,0)$ transforms as
\begin{equation}
    \mathbf b_s\longmapsto R_z(\theta)\mathbf b_s.
    \label{eq:classical_b_source_rotation}
\end{equation}
Consequently,
\begin{equation}
    \left[R_z(\theta)\mathbf L_h\right]\cdot
    \left[R_z(\theta)\mathbf b_s\right]
    =\mathbf L_h\cdot\mathbf b_s,
    \label{eq:classical_interaction_rotation_invariance}
\end{equation}
and the interaction Hamiltonian is invariant.  The free terms are also
invariant because they depend only on $\mathbf L_h^2$, $L^z_h$, and
$X^2+P^2$.  Hence the complete Hamiltonian in
Eq.~\eqref{eq:classical_joint_hamiltonian} is invariant under the joint
rotation.

The infinitesimal generator of this transformation is the additive observable
\begin{equation}
    Q=Q_h+Q_s=L^z_h+J_s
    =L^z_h+\frac{X^2+P^2}{2}.
    \label{eq:classical_total_charge}
\end{equation}
To see this directly, let $\theta$ play the role of the transformation
parameter.  For the harvester,
\begin{align}
    \frac{dL^x_h}{d\theta}
    &=\{L^x_h,Q\}=-L^y_h,
    &
    \frac{dL^y_h}{d\theta}
    &=\{L^y_h,Q\}=L^x_h,
    &
    \frac{dL^z_h}{d\theta}
    &=0,
    \label{eq:classical_Q_generates_L_rotation}
\end{align}
which generates $R_z(\theta)$.  For the source,
\begin{equation}
    \frac{dX}{d\theta}=\{X,Q\}=P,
    \qquad
    \frac{dP}{d\theta}=\{P,Q\}=-X,
    \label{eq:classical_Q_generates_source_rotation}
\end{equation}
which generates Eq.~\eqref{eq:classical_source_rotation}.

We may also verify the conservation law explicitly.  The free Hamiltonians
Poisson commute with $Q$.  For the interaction term,
\begin{align}
    \{H_{\rm int},L^z_h\}_h
    &=g(XL^y_h+PL^x_h),
    \label{eq:classical_Hint_Lz_bracket}\\
    \{H_{\rm int},J_s\}_s
    &=-g(PL^x_h+XL^y_h).
    \label{eq:classical_Hint_J_bracket}
\end{align}
The two contributions cancel, giving
\begin{equation}
    \boxed{\{H,Q\}=0.}
    \label{eq:classical_noether_relation}
\end{equation}
Equivalently,
\begin{equation}
    \frac{dQ}{dt}=0.
    \label{eq:classical_Q_conserved}
\end{equation}

\subsection{Reduced classical dynamics and the phase orbit}
\label{app:classical_reduced_covariance}

For completeness, we now express the symmetry statement at the level of
classical states.  Let $f_h(\mathbf L_h)$ and $f_s(X,P)$ be normalized initial
phase-space densities, and assume the initial joint state is their product.
Let $\mathcal T_T$ denote the Liouville evolution generated by the joint
Hamiltonian for a time $T$.  For a fixed source state $f_s$, the induced
harvester map is
\begin{equation}
    \Phi_{f_s}^{T}[f_h](\mathbf L_h)
    :=\int dX\,dP\,
      \mathcal T_T[f_hf_s](\mathbf L_h,X,P).
    \label{eq:classical_reduced_map_specific}
\end{equation}

Let $\mathcal R_h(\theta)$ and $\mathcal R_s(\theta)$ denote the induced
actions of Eqs.~\eqref{eq:classical_harvester_rotation} and
\eqref{eq:classical_source_rotation} on phase-space densities.  Since
$\{H,Q\}=0$, the joint rotation commutes with the Hamiltonian evolution.  The
reduced map consequently satisfies
\begin{equation}
    \Phi_{\mathcal R_s(\theta)f_s}^{T}[f_h]
    =\mathcal R_h(\theta)
      \Phi_{f_s}^{T}
      [\mathcal R_h(-\theta)f_h].
    \label{eq:classical_specific_covariance}
\end{equation}
For a source localized at action $J_0$ and phase $\phi_0$, one may write
\begin{equation}
    f_{s,\phi_0}(X,P)
    =\delta\!\left(X-\sqrt{2J_0}\cos\phi_0\right)
     \delta\!\left(P+\sqrt{2J_0}\sin\phi_0\right).
    \label{eq:classical_source_point_state}
\end{equation}
The source transformation shifts its phase,
\begin{equation}
    \mathcal R_s(\theta)f_{s,\phi_0}
    =f_{s,\phi_0+\theta}.
    \label{eq:classical_source_phase_shift}
\end{equation}

\noindent We take the initial and final harvester states to be concentrated at opposite
poles,
\begin{equation}
    \mathbf L_{h,i}=-L_h\hat{\mathbf z},
    \qquad
    \mathbf L_{h,f}=+L_h\hat{\mathbf z}.
    \label{eq:classical_boundary_vectors}
\end{equation}
The corresponding point measures, denoted by $f_{h,i}$ and $f_{h,f}$, are
invariant under every rotation about $z$:
\begin{equation}
    \mathcal R_h(\theta)f_{h,i}=f_{h,i},
    \qquad
    \mathcal R_h(\theta)f_{h,f}=f_{h,f}.
    \label{eq:classical_boundary_invariance}
\end{equation}
Suppose that a seed source phase $\phi_0$ achieves the deterministic
transition
\begin{equation}
    \Phi_{f_{s,\phi_0}}^{T}[f_{h,i}]=f_{h,f}.
    \label{eq:classical_seed_transition_assumed}
\end{equation}
Using Eqs.~\eqref{eq:classical_specific_covariance},
\eqref{eq:classical_source_phase_shift}, and
\eqref{eq:classical_boundary_invariance}, we find
\begin{align}
    \Phi_{f_{s,\phi_0+\theta}}^{T}[f_{h,i}]
    &=\mathcal R_h(\theta)
      \Phi_{f_{s,\phi_0}}^{T}
      [\mathcal R_h(-\theta)f_{h,i}] \nonumber  \\
    &=\mathcal R_h(\theta)
      \Phi_{f_{s,\phi_0}}^{T}[f_{h,i}] \nonumber \\
    &=\mathcal R_h(\theta)f_{h,f} \nonumber \\
    &=f_{h,f}.
    \label{eq:classical_phase_family_transition}
\end{align}
Thus every source phase at the same source action $J_0$ produces the same
harvester transition at the same time.  We next show explicitly that such a
seed trajectory exists for the Hamiltonian in
Eq.~\eqref{eq:classical_joint_hamiltonian}.

\subsection{Explicit deterministic energy-harvesting seed}
\label{app:classical_deh_seed}

We now set
\begin{equation}
    \Omega=\omega,
    \label{eq:classical_resonance_condition}
\end{equation}
and choose the initial conditions
\begin{equation}
    \mathbf L_h(0)=-L_h\hat{\mathbf z},
    \qquad
    J_s(0)=J_0,
    \label{eq:classical_seed_initial_LJ}
\end{equation}
with
\begin{equation}
    X(0)=\sqrt{2J_0}\cos\phi_0,
    \qquad
    P(0)=-\sqrt{2J_0}\sin\phi_0.
    \label{eq:classical_seed_initial_XP}
\end{equation}
Because $L^x_h(0)=L^y_h(0)=0$, the initial interaction energy vanishes:
\begin{equation}
    H_{\rm int}(0)=0.
    \label{eq:classical_seed_initial_Hint_zero}
\end{equation}

At $\Omega=\omega$, the total Hamiltonian can be rewritten as
\begin{align}
    H
    &=\frac{\mathbf L_h^2}{2\mathfrak I_h}
      +\omega L^z_h+\omega J_s+H_{\rm int} \\
    &=\frac{L_h^2}{2\mathfrak I_h}+\omega Q+H_{\rm int}.
    \label{eq:classical_resonant_H_Q}
\end{align}
The quantities $H$, $L_h$, and $Q$ are all conserved.  It follows from
Eq.~\eqref{eq:classical_resonant_H_Q} that $H_{\rm int}$ is itself constant
along this resonant trajectory.  Together with
Eq.~\eqref{eq:classical_seed_initial_Hint_zero}, this gives
\begin{equation}
    H_{\rm int}(t)=0
    \qquad\text{for all }t.
    \label{eq:classical_Hint_zero_trajectory}
\end{equation}
For $g\neq0$, Eq.~\eqref{eq:classical_Hint_zero_trajectory} is equivalent to
\begin{equation}
    XL^x_h-PL^y_h=0.
    \label{eq:classical_seed_orthogonality_constraint}
\end{equation}

Conservation of $Q=L^z_h+J_s$ gives
\begin{equation}
    J_s=Q-L^z_h,
    \qquad
    X^2+P^2=2(Q-L^z_h).
    \label{eq:classical_source_action_vs_z}
\end{equation}
Conservation of $|\mathbf L_h|=L_h$ similarly gives
\begin{equation}
    (L^x_h)^2+(L^y_h)^2=L_h^2-(L^z_h)^2.
    \label{eq:classical_transverse_L_vs_z}
\end{equation}
From Eq.~\eqref{eq:classical_joint_Lz_eom},
\begin{equation}
    \dot L^z_h=-g(PL^x_h+XL^y_h).
    \label{eq:classical_seed_z_dot}
\end{equation}
The algebraic identity
\begin{equation}
    (XL^x_h-PL^y_h)^2+(PL^x_h+XL^y_h)^2
    =(X^2+P^2)\left[(L^x_h)^2+(L^y_h)^2\right]
    \label{eq:classical_seed_algebraic_identity}
\end{equation}
can be verified by expanding both sides. The first term on the left vanishes
by Eq.~\eqref{eq:classical_seed_orthogonality_constraint}. Substituting
Eqs.~\eqref{eq:classical_source_action_vs_z} and
\eqref{eq:classical_transverse_L_vs_z} therefore gives
\begin{equation}
    (PL^x_h+XL^y_h)^2
    =2(Q-L^z_h)\left[L_h^2-(L^z_h)^2\right].
    \label{eq:classical_seed_combination_squared}
\end{equation}
Squaring Eq.~\eqref{eq:classical_seed_z_dot}, we arrive at the closed equation
\begin{equation}
    \boxed{
    \left(\dot L^z_h\right)^2
    =2g^2(Q-L^z_h)\left[L_h^2-(L^z_h)^2\right].}
    \label{eq:classical_seed_z_closed}
\end{equation}

Although $\dot L^z_h(0)=0$ at the south pole, the trajectory initially moves
towards increasing $L^z_h$. To verify this explicitly, define
\begin{equation}
    C:=PL^x_h+XL^y_h,
    \qquad
    \dot L^z_h=-gC.
    \label{eq:classical_C_definition}
\end{equation}
At $t=0$, $L^x_h=L^y_h=0$, and Eqs.~\eqref{eq:classical_joint_Lx_eom} and
\eqref{eq:classical_joint_Ly_eom} give
\begin{equation}
    \dot L^x_h(0)=-gL_h P(0),
    \qquad
    \dot L^y_h(0)=-gL_h X(0).
    \label{eq:classical_initial_transverse_derivatives}
\end{equation}
Therefore,
\begin{align}
    \dot C(0)
    &=P(0)\dot L^x_h(0)+X(0)\dot L^y_h(0) \nonumber \\
    &=-gL_h\left[P(0)^2+X(0)^2\right] \nonumber \\
    &=-2gL_h J_0,
    \label{eq:classical_C_dot_initial}
\end{align}
and hence
\begin{equation}
    \ddot L^z_h(0)=-g\dot C(0)=2g^2L_h J_0>0.
    \label{eq:classical_z_acceleration_initial}
\end{equation}
The increasing branch of Eq.~\eqref{eq:classical_seed_z_closed} is therefore
\begin{equation}
    \dot L^z_h
    =|g|\sqrt{2(Q-L^z_h)\left[L_h^2-(L^z_h)^2\right]}.
    \label{eq:classical_seed_z_increasing}
\end{equation}

For the initial state in Eq.~\eqref{eq:classical_seed_initial_LJ}, the
conserved charge is
\begin{equation}
    Q=J_0-L_h.
    \label{eq:classical_seed_Q_initial}
\end{equation}
If the trajectory reaches $L^z_h=+L_h$, the final source action must be
\begin{equation}
    J_{s,f}=Q-L_h=J_0-2L_h.
    \label{eq:classical_seed_final_action}
\end{equation}
Thus $J_0\geq2L_h$ is required for nonnegative final source action. The
strict condition
\begin{equation}
    J_0>2L_h,
    \label{eq:classical_seed_action_condition}
\end{equation}
or equivalently $Q>L_h$, ensures that the pole is reached in finite time.
Integrating Eq.~\eqref{eq:classical_seed_z_increasing} gives
\begin{equation}
    t(L^z_h)
    =\frac{1}{|g|}
      \int_{-L_h}^{L^z_h}
      \frac{du}{\sqrt{2(Q-u)(L_h^2-u^2)}}.
    \label{eq:classical_seed_time_to_z}
\end{equation}
The complete transfer time is therefore
\begin{equation}
    \boxed{
    T(J_0)
    =\frac{1}{|g|}
      \int_{-L_h}^{+L_h}
      \frac{du}
      {\sqrt{2(J_0-L_h-u)(L_h^2-u^2)}}.}
    \label{eq:classical_seed_transfer_time}
\end{equation}
The integral is independent of the initial source phase $\phi_0$. Hence the
same stopping time applies to the entire phase orbit in
Eq.~\eqref{eq:classical_phase_family_transition}.

In the large-source limit $J_0\gg L_h$, the source action changes only by a
small fraction during the transfer. Approximating
$J_0-L_h-u\simeq J_0$ throughout the integration range in
Eq.~\eqref{eq:classical_seed_transfer_time} gives
\begin{align}
    T(J_0)
    &\simeq\frac{1}{|g|\sqrt{2J_0}}
      \int_{-L_h}^{+L_h}
      \frac{du}{\sqrt{L_h^2-u^2}} \nonumber \\
    &=\frac{\pi}{|g|\sqrt{2J_0}}.
    \label{eq:classical_seed_large_source_time}
\end{align}
Since the source-orbit amplitude is $A=\sqrt{2J_0}$ and
$|g|A=|\beta|B_1$, this reduces to
\begin{equation}
    T\simeq\frac{\pi}{|g|A}
    =\frac{\pi}{|\beta|B_1},
    \label{eq:classical_seed_external_limit_time}
\end{equation}
which is the half-rotation time generated by an effectively prescribed
transverse field of magnitude $B_1$.

Finally, the energy bookkeeping is explicit. Because the interaction
vanishes at both poles, the harvester energy change is
\begin{align}
    W_h
    &=H_h(+L_h\hat{\mathbf z})
      -H_h(-L_h\hat{\mathbf z}) \nonumber \\
    &=2\Omega L_h.
    \label{eq:classical_seed_harvester_energy}
\end{align}
The source action decreases by $2L_h$, so
\begin{equation}
    \Delta H_s
    =\omega(J_{s,f}-J_0)
    =-2\omega L_h.
    \label{eq:classical_seed_source_energy}
\end{equation}
At $\Omega=\omega$,
\begin{equation}
    \Delta H_s+W_h=0,
    \label{eq:classical_seed_energy_balance}
\end{equation}
as required by conservation of the autonomous joint Hamiltonian. For
$\Omega>0$, the transition
$-L_h\hat{\mathbf z}\rightarrow+L_h\hat{\mathbf z}$ increases the
harvester energy. If $\Omega<0$, the charging direction is reversed.

Equations~\eqref{eq:classical_phase_family_transition} and
\eqref{eq:classical_seed_transfer_time} together establish a classical DEH
family: every source state with the same initial action $J_0$ and arbitrary
phase $\phi_0$ drives the harvester between the same two boundary states at
the same time $T(J_0)$, while transferring the same amount of energy $W_h$.

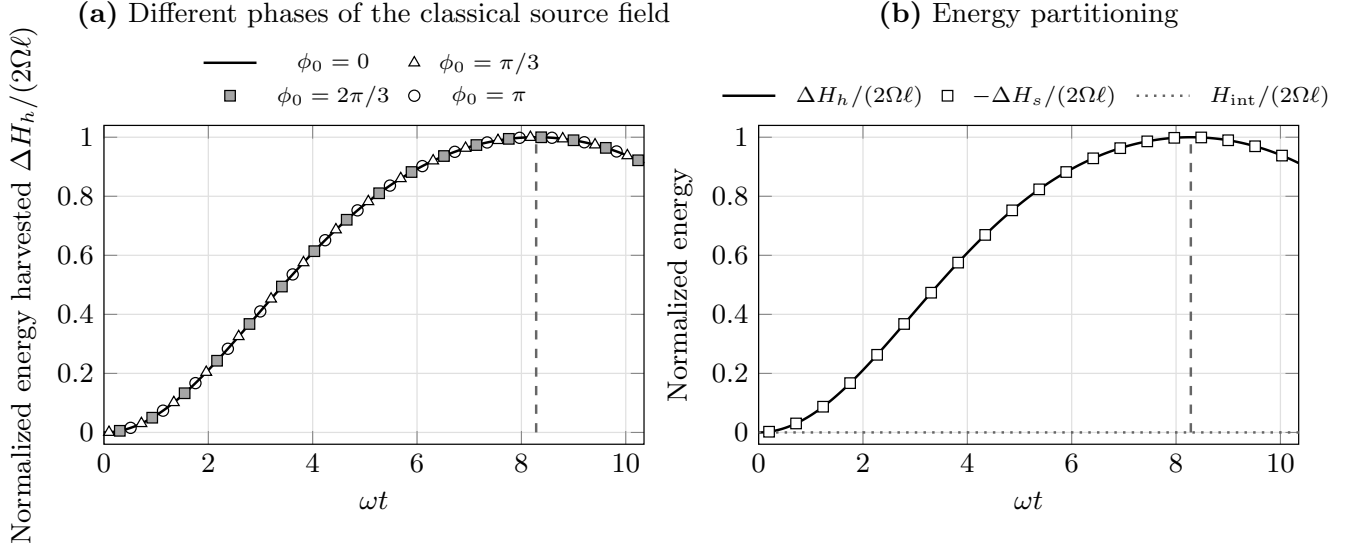
\begin{figure*}[t]
    \centering
\resizebox{\linewidth}{!}{%
\begin{tikzpicture}
\begin{groupplot}[
    group style={
        group size=2 by 1,
        horizontal sep=1.30cm,
    },
    width=0.43\linewidth,
    height=0.29\linewidth,
    xmin=0,
    xmax=10.35,
    ymin=-0.04,
    ymax=1.04,
    xlabel={\(\omega t\)},
    xtick={0,2,4,6,8,10},
    ytick={0,0.2,0.4,0.6,0.8,1.0},
    tick label style={font=\small},
    label style={font=\small},
    title style={font=\small, yshift=2.35em},
    grid=major,
    grid style={black!12},
    legend style={
        draw=none,
        fill=none,
        font=\scriptsize,
        column sep=0.45em,
    },
]

\nextgroupplot[
    title={\textbf{(a)} Different phases of the classical source field},
    ylabel={Normalized energy harvested
        \(\Delta H_h/(2\Omega\ell)\)},
    legend style={
        at={(0.5,1.02)},
        anchor=south,
        legend columns=2,
        draw=none,
        fill=none,
        font=\scriptsize,
        column sep=0.45em,
        row sep=0pt,
    },
]

\addplot[
    black,
    thick,
]
table [x=omega_t, y=gain_phi_0, col sep=space]
{classical_deh_phase_orbit.dat};
\addlegendentry{\(\phi_0=0\)}

\addplot[
    only marks,
    mark=triangle*,
    mark size=2.0pt,
    mark options={draw=black, fill=white},
    mark repeat=90,
    mark phase=15,
]
table [x=omega_t, y=gain_phi_pi_over_3, col sep=space]
{classical_deh_phase_orbit.dat};
\addlegendentry{\(\phi_0=\pi/3\)}

\addplot[
    only marks,
    mark=square*,
    mark size=1.7pt,
    mark options={draw=black, fill=black!35},
    mark repeat=90,
    mark phase=45,
]
table [x=omega_t, y=gain_phi_2pi_over_3, col sep=space]
{classical_deh_phase_orbit.dat};
\addlegendentry{\(\phi_0=2\pi/3\)}

\addplot[
    only marks,
    mark=o,
    mark size=1.8pt,
    mark options={draw=black},
    mark repeat=90,
    mark phase=75,
]
table [x=omega_t, y=gain_phi_pi, col sep=space]
{classical_deh_phase_orbit.dat};
\addlegendentry{\(\phi_0=\pi\)}

\addplot[
    dashed,
    black!60,
    thick,
    forget plot,
]
table [x=omega_t, y=normalized_energy, col sep=space]
{classical_deh_transfer_time.dat};

\nextgroupplot[
    title={\textbf{(b)} Energy partitioning},
    ylabel={Normalized energy},
    legend style={
        at={(0.5,1.02)},
        anchor=south,
        legend columns=3,
        draw=none,
        fill=none,
        font=\scriptsize,
        column sep=0.45em,
    },
]

\addplot[
    black,
    thick,
]
table [x=omega_t, y=harvester_gain, col sep=space]
{classical_deh_energy_balance.dat};
\addlegendentry{\(\Delta H_h/(2\Omega\ell)\)}

\addplot[
    only marks,
    mark=square*,
    mark size=1.7pt,
    mark options={draw=black, fill=white},
    mark repeat=75,
    mark phase=30,
]
table [x=omega_t, y=source_loss, col sep=space]
{classical_deh_energy_balance.dat};
\addlegendentry{\(-\Delta H_s/(2\Omega\ell)\)}

\addplot[
    black!60,
    thick,
    dotted,
]
table [x=omega_t, y=interaction_energy, col sep=space]
{classical_deh_energy_balance.dat};
\addlegendentry{\(H_{\mathrm{int}}/(2\Omega\ell)\)}

\addplot[
    dashed,
    black!60,
    thick,
    forget plot,
]
table [x=omega_t, y=normalized_energy, col sep=space]
{classical_deh_transfer_time.dat};

\end{groupplot}
\end{tikzpicture}
}%
    \caption{\justifying
    Classical deterministic energy harvesting.
    (a) Normalized harvester energy gain for different initial phases
    $\phi_0$ of the classical source field. All phase-dependent results
    coincide and reach unity at the common transfer time $T$, indicated
    by the vertical dashed line.
    (b) Corresponding energy partitioning for a representative initial
    phase. The harvester energy gain equals the source energy loss, while
    the interaction energy remains zero along the selected seed source state trajectory.}
    \label{fig:classical_DEH}
\end{figure*}

\section{Asymmetry inequality and equality for exact DEH}
\label{sec:asymmetry_SM}

Here we show that a symmetry-preserving harvesting process cannot increase source asymmetry on average, and that this inequality is saturated for exact DEH with pure harvester boundary states.

\subsection{Asymmetry inequality}

Consider a symmetry-invariant initial harvester state \(\rho_h\) and a symmetry-invariant projective measurement \(\{\Pi_k\}\) performed on the harvester after the interaction. The conditional transformations of the source are described by the quantum instrument
\begin{equation}
\Lambda_\rho^k(X)
:=
\Tr_h\!\left[
(\Pi_k\otimes I_s)
U(\tau)(\rho_h\otimes X)U^\dagger(\tau)
(\Pi_k\otimes I_s)
\right].
\label{eq:SM_instrument}
\end{equation}
The operations \(\{\Lambda_\rho^k\}_k\) are completely positive and trace non-increasing, and their sum is trace preserving. They induce on the source the POVM
\begin{equation}
M_s(k):=(\Lambda_\rho^k)^\dagger(I_s),
\end{equation}
so that the probability of outcome \(k\) for a source state \(\sigma\) is
\begin{equation}
p_k=\Tr[M_s(k)\sigma]
=\Tr[\Lambda_\rho^k(\sigma)].
\end{equation}

The symmetry of the global dynamics makes every conditional source operation covariant. Indeed, using
\begin{equation}
[U(\tau),R_h(\mathfrak g)\otimes R_s(\mathfrak g)]=0,
\end{equation}
together with
\begin{equation}
R_h(\mathfrak g)\rho_hR_h^\dagger(\mathfrak g)=\rho_h,
\qquad
R_h(\mathfrak g)\Pi_kR_h^\dagger(\mathfrak g)=\Pi_k,
\end{equation}
in Eq.~\eqref{eq:SM_instrument} gives
\begin{equation}
\Lambda_\rho^k\!\left(
R_s(\mathfrak g)XR_s^\dagger(\mathfrak g)
\right)
=
R_s(\mathfrak g)\Lambda_\rho^k(X)R_s^\dagger(\mathfrak g).
\label{eq:SM_instrument_cov}
\end{equation}
Consequently each \(M_s(k)\) is symmetry invariant.

Let \(\mathcal I\) be an asymmetry monotone satisfying selective monotonicity under covariant operations \cite{Takagi2018}. Defining the normalized conditional source state
\begin{equation}
\sigma_k:=\frac{\Lambda_\rho^k(\sigma)}{p_k}
\end{equation}
for \(p_k>0\), Eq.~\eqref{eq:SM_instrument_cov} implies
\begin{equation}
\boxed{
\mathcal I(\sigma)
\geq
\sum_k p_k\,\mathcal I(\sigma_k).
}
\label{eq:SM_asymmetry_inequality}
\end{equation}
Thus symmetry-preserving harvesting cannot increase the source asymmetry on average, although the asymmetry conditioned on an individual outcome may increase.

\subsection{Exact DEH: saturation of the inequality}

For exact DEH with pure harvester boundary states, the asymmetry inequality is saturated. Let the harvester initially be in \(|g\rangle_h\) and end deterministically in \(|e\rangle_h\), and define
\begin{equation}
A_{eg}(\tau):={}_h\langle e|U(\tau)|g\rangle_h .
\label{eq:SM_Ke}
\end{equation}
The final source state is then
\begin{equation}
\sigma'=A_{eg}(\tau)\sigma A_{eg}(\tau)^\dagger.
\label{eq:SM_sigma_prime}
\end{equation}
Since \(A_{eg}(\tau)\) is a Kraus operator obtained from a unitary interaction,
\begin{equation}
A_{eg}(\tau)^\dagger A_{eg}(\tau)\leq I_s.
\end{equation}
Exact DEH implies that the outcome \(e\) has probability one,
\begin{equation}
1=\Tr[A_{eg}(\tau)^\dagger A_{eg}(\tau)\,\sigma].
\end{equation}
Writing \(P\) for the projector onto \(\operatorname{supp}\sigma\), positivity of
\(I_s-A_{eg}(\tau)^\dagger A_{eg}(\tau)\) therefore gives
\begin{equation}
A_{eg}(\tau)^\dagger A_{eg}(\tau) P=P.
\label{eq:SM_isometry}
\end{equation}
Hence \(A_{eg}(\tau)\) is an isometry on the support of the DEH source state. In particular, \(\sigma\) and \(\sigma'\) have the same nonzero spectrum, and therefore
\begin{equation}
S(\sigma')=S(\sigma).
\label{eq:SM_entropy}
\end{equation}

The isometry also respects the symmetry in the appropriate sense. Since the pure boundary states are symmetry invariant as density operators, there exist phases \(\alpha_g(\mathfrak g)\) and \(\alpha_e(\mathfrak g)\) such that
\begin{equation}
R_h(\mathfrak g)|g\rangle
=e^{i\alpha_g(\mathfrak g)}|g\rangle,
\qquad
R_h(\mathfrak g)|e\rangle
=e^{i\alpha_e(\mathfrak g)}|e\rangle .
\end{equation}
Taking the \({}_h\langle e|\,\cdot\,|g\rangle_h\) matrix element of the global symmetry relation gives
\begin{equation}
A_{eg}(\tau)R_s(\mathfrak g)
=
e^{i\vartheta(\mathfrak g)}
R_s(\mathfrak g)A_{eg}(\tau),
\qquad
\vartheta(\mathfrak g)
=
\alpha_e(\mathfrak g)-\alpha_g(\mathfrak g).
\label{eq:SM_intertwining}
\end{equation}
Thus \(A_{eg}(\tau)\) need not commute with the source charge; it may carry a definite charge shift. Nevertheless, the phase in Eq.~\eqref{eq:SM_intertwining} cancels in the operation \(A_{eg}(\tau)(\cdot)A_{eg}(\tau)^\dagger\). Moreover, Eq.~\eqref{eq:SM_intertwining} implies that both \(A_{eg}(\tau)^\dagger A_{eg}(\tau)\) and \(A_{eg}(\tau)A_{eg}(\tau)^\dagger\) commute with \(R_s(\mathfrak g)\).

To establish equality for any asymmetry monotone, we extend the transformation in Eq.~\eqref{eq:SM_sigma_prime} to covariant channels in both directions. Define
\begin{align}
F&:=\sqrt{I_s-A_{eg}(\tau)^\dagger A_{eg}(\tau)},\\
\mathcal C_{\rightarrow}(X)
&:=
A_{eg}(\tau)XA_{eg}(\tau)^\dagger+FXF.
\label{eq:SM_forward_channel}
\end{align}
Since \(A_{eg}(\tau)^\dagger A_{eg}(\tau)+F^2=I_s\), \(\mathcal C_{\rightarrow}\) is trace preserving. Furthermore, \(F\) commutes with the symmetry, while \(A_{eg}(\tau)\) transforms according to Eq.~\eqref{eq:SM_intertwining}; hence \(\mathcal C_{\rightarrow}\) is covariant. Equation~\eqref{eq:SM_isometry} implies \(FP=0\), and therefore
\begin{equation}
\mathcal C_{\rightarrow}(\sigma)
=
A_{eg}(\tau)\sigma A_{eg}(\tau)^\dagger
=
\sigma'.
\label{eq:SM_forward}
\end{equation}

A covariant reverse channel can be constructed similarly. Define
\begin{align}
G&:=\sqrt{I_s-A_{eg}(\tau)A_{eg}(\tau)^\dagger},\\
\mathcal C_{\leftarrow}(X)
&:=
A_{eg}(\tau)^\dagger X A_{eg}(\tau)+GXG.
\label{eq:SM_reverse_channel}
\end{align}
Again, \(\mathcal C_{\leftarrow}\) is trace preserving and covariant. To see that it reverses the transformation on \(\sigma'\), first note that
\begin{equation}
(I_s-A_{eg}(\tau)A_{eg}(\tau)^\dagger)A_{eg}(\tau)P
=
A_{eg}(\tau)(I_s-A_{eg}(\tau)^\dagger A_{eg}(\tau))P
=
0.
\end{equation}
Since \(G\) is the positive square root of \(I_s-A_{eg}(\tau)A_{eg}(\tau)^\dagger\), it has the same kernel, and hence
\begin{equation}
GA_{eg}(\tau)P=0.
\label{eq:SM_Gannihilates}
\end{equation}
Because \(\operatorname{supp}\sigma'\subseteq\operatorname{Ran}(A_{eg}(\tau)P)\), Eq.~\eqref{eq:SM_Gannihilates} gives \(G\sigma'G=0\). Using Eq.~\eqref{eq:SM_isometry}, we therefore obtain
\begin{align}
\mathcal C_{\leftarrow}(\sigma')
&=
A_{eg}(\tau)^\dagger\sigma'A_{eg}(\tau)
\nonumber\\
&=
A_{eg}(\tau)^\dagger A_{eg}(\tau)\,\sigma\,A_{eg}(\tau)^\dagger A_{eg}(\tau)
=
\sigma.
\label{eq:SM_reverse}
\end{align}

The initial and final source states are therefore reversibly connected by covariant channels,
\begin{equation}
\sigma
\xrightarrow{\ \mathcal C_{\rightarrow}\ }
\sigma'
\xrightarrow{\ \mathcal C_{\leftarrow}\ }
\sigma .
\end{equation}
For any asymmetry monotone \(\mathcal I\) that is non-increasing under covariant channels,
\begin{equation}
\mathcal I(\sigma)
\geq
\mathcal I(\sigma')
\geq
\mathcal I(\sigma),
\end{equation}
and hence
\begin{equation}
\boxed{
\mathcal I(\sigma')
=
\mathcal I(\sigma).
}
\label{eq:SM_asymmetry_equality}
\end{equation}
Thus exact DEH transfers energy from the source while preserving both its entropy and its asymmetry. By the Noether-DEH theorem, the same conclusion holds for every source state on the symmetry orbit of a DEH seed. Since mixtures of DEH-capable source states are also DEH capable, Eq.~\eqref{eq:SM_asymmetry_equality} likewise applies to any mixture over such an orbit.

\bibliographystyle{apsrev4-2}